\documentclass[usenatbib,usegraphicx,floats]{mn2e}
\usepackage{natbib}
\bibpunct{(}{)}{;}{a}{}{,}
\usepackage{graphicx}
\usepackage{psfig}

\begin{document}

\title{The runaway instability of thick discs around black
holes.\\
I. The constant angular momentum case}

\author[Font \& Daigne]
{Jos\'{e} A. Font$^{1,2}$
and Fr\'{e}d\'{e}ric Daigne$^{1,3}$\\
$^{1}$ Max-Planck-Institut f\"{u}r Astrophysik, Karl-Schwarzschild-Str. 1,
81748 Garching bei M\"{u}nchen, Germany\\
$^{2}$ Present address: Departamento de Astronom\'{\i}a y Astrof\'{\i}sica,
Universidad de Valencia, 46100 Burjassot (Valencia), Spain\\
$^{3}$ Present address: CEA/DSM/DAPNIA, Service d'Astrophysique, C.E. Saclay, 
91191 Gif sur Yvette Cedex, France\\
}
\maketitle

\begin{abstract}
We present results from a numerical study of the runaway instability of
thick discs around black holes. This instability is an important issue for 
most models of cosmic gamma-ray bursts, where the central engine responsible 
for the initial energy release is such a system consisting of a thick disc
surrounding a black hole.  We have carried out a comprehensive number
of time-dependent simulations aimed at exploring the appearance of the
instability. Our study has been performed using a fully relativistic
hydrodynamics code. The general relativistic hydrodynamic equations 
are formulated as a hyperbolic flux-conservative system and solved using 
a suitable Godunov-type scheme. We build a series of constant angular
momentum discs around a Schwarzschild black hole. Furthermore,
the self-gravity of the disc is neglected and the evolution of the central
black hole is assumed to be that of a sequence of exact Schwarzschild black
holes of varying mass. The black hole mass increase is thus determined by
the mass accretion rate across the event horizon. In agreement with
previous studies based on stationary models, we find that by allowing the
mass of the black hole to grow the disc becomes unstable. Our hydrodynamical
simulations show that for all disc-to-hole mass ratios considered
(between 1 and 0.05), the runaway instability appears very fast on a
dynamical timescale of a few orbital periods, typically a few 10 ms and
never exceeding 1 s for our particular choice of the mass of the black hole 
($2.5\ \mathrm{M_\odot}$) and a large range of mass fluxes ($\dot{m} 
\ga 10^{-3}\ \mathrm{M_{\odot}/s}$). The implications of our results in 
the context of gamma-ray bursts are briefly discussed.
\end{abstract}

\begin{keywords}
accretion, accretion discs -- black hole physics -- hydrodynamics -- 
instabilities -- gamma rays: bursts.
\end{keywords}


\section{Introduction}

\begin{table*}
\caption{Summary of representative results concerning the runaway instability
on \textit{stationary} models of thick discs around black holes.}
\begin{tabular}{lllllll|}
\hline
Reference & Framework & Angular momentum & BH rotation &
Disc self-gravity & Results\\
\hline
\hline
\citet{abramowicz:83} & Pseudo-Newt. & $l=\mathrm{constant}$
& no & approximate & \begin{minipage}[t]{3.5cm}
unstable$^{1}$
\end{minipage}\\
\citet{wilson:84}     & GR     & $l=\mathrm{constant}$ & yes & neglected & stable \\
\citet{khanna:92}     & Pseudo-Newt. & $l=\mathrm{constant}$ & yes & approximate & unstable \\
\citet{nishida:96a}   & GR           & $l=\mathrm{constant}$ & yes & exact & unstable$^{2}$ \\
\citet{daigne:97}     & Pseudo-Newt. & $l \propto \varpi^{\alpha}$, $0 \le \alpha \le 0.5$ & no 
& neglected & stable$^{3}$ \\
\citet{abramowicz:98} & GR           & $l \propto \varpi^{\alpha}$, $0 \le \alpha \le 0.5$ & yes 
& neglected & stable$^{4}$\\
\citet{masuda:98}     & Pseudo-Newt. & $l \propto \varpi^{\alpha}$, $\alpha=0.2$ & no 
& exact & unstable$^{5}$ \\
\citet{lu:00}         & Pseudo-Newt. & $l \propto \varpi^{\alpha}$, $0 \le \alpha \le 0.5$ & no 
& neglected & stable$^{6}$\\
\hline
\end{tabular}
\begin{tabular}{l}
Notes :\\
$^{1}$ for a large range of disc-to-hole mass ratio and disc inner radius.\\
$^{2}$ study made for only four intial models.\\
$^{3}$ The mass of the black hole is $2.44\ \mathrm{M}_\odot$. The mass of the disc is 
$0.36\ \mathrm{M}_\odot$. The system is stable for $\alpha \ge \alpha_\mathrm{cr}$ with 
$\alpha_\mathrm{cr} \sim 0.1$.\\
$^{4}$ Same parameters as \citet{daigne:97}. The critical value $\alpha_\mathrm{cr}$ 
decreases when the black hole rotation increases.\\
$^{5}$ Same parameters as \citet{daigne:97}. The system becomes unstable for large 
transfer of mass only.\\
$^{6}$ Same conclusion as \citet{daigne:97} for a completely different range of masses 
(massive black hole of mass $10^{6}\ \mathrm{M}_\odot$).
\end{tabular}
\label{tab:studies}
\end{table*}

%
%

Thick accretion discs are probably present in many astrophysical objects,
e.g. quasars and other active galactic nuclei, some X-ray binaries and microquasars, 
and the central engine of gamma-ray bursts (GRBs hereafter). They have
been studied in great detail by many authors (see e.g. 
\citet{rees:84} and references therein).
In particular, it is well known that in a system formed
by a black hole surrounded by a thick disc, the gas flows in an effective
(gravitational plus centrifugal)  potential, whose structure is comparable
with that of a close binary. The Roche torus surrounding the black hole
has a cusp-like inner edge located at the Lagrange point $\mathrm{L}_{1}$
where mass transfer driven by the radial pressure gradient is possible
\citep{abramowicz:78}.

The so-called runaway instability of such systems was first discovered
by \citet{abramowicz:83}. The underlying mechanism is the following:
due to accretion of material from the disc, the mass of the black hole
increases and the gravitational field of the system changes. Therefore,
an accretion disc can never reach a completely steady state. Starting
from an initial disc filling its Roche lobe so that mass transfer is
possible through the cusp located at the $\mathrm{L}_{1}$ Lagrange point,
two evolutions are feasible when the mass of the black hole increases:
(i) either the cusp moves inwards towards the black hole, which slows
down the mass transfer, resulting in a stable situation, or (ii) the
cusp moves deeper inside the disc material. In this case the mass transfer
speeds up, leading to the runaway instability.

%
%

In their first study, \citet{abramowicz:83} analyzed the effect of the
mass transfer under many simplifying assumptions (a pseudo-Newtonian
potential for the black hole \citep{paczynski:80}, constant angular
momentum in the disc, and approximate treatment of the disc self-gravity).
These authors found that the runaway instability occurs for a large range
of parameters such as the disc-to-hole mass ratio and the location of
the disc inner radius. More detailed studies followed, whose main
conclusions are summarized in Table \ref{tab:studies}. Notice that all
these studies are based on stationary models in which a fraction of the
mass and angular momentum of an initial disc filling its Roche lobe
is transfered to the black hole, and the new gravitational field
is used to compute the new position of the cusp, which controls whether
the runaway instability occurs or not. The conclusions of these studies
have yet to be confirmed on a dynamical framework.

From Table \ref{tab:studies} one sees that (i) taking into account
the self-gravity of the disc seems to favor the instability
\citep{khanna:92,nishida:96a,masuda:98}; (ii) the rotation of the
black hole has a stabilizing effect \citep{wilson:84,abramowicz:98};
(iii) taking into account a non-constant distribution of the angular
momentum (increasing outwards) has a {\it strong} stabilizing effect
\citep{daigne:97,abramowicz:98};  (iv) using a fully relativistic
potential instead of a pseudo-Newtonian potential for the black hole
seems to act in the direction of favoring the instability
\citep{nishida:96a}. It also becomes evident from Table \ref{tab:studies}
that there is not still a final consensus about the very existence of
the instability. In the fully relativistic study of \citet{nishida:96a},
it was shown that the runaway instability occurs when the angular
momentum is constant in the disc. However, the work of \citet{daigne:97}
showed the stabilizing effect of a distribution of angular momentum
in the disc increasing outwards. It is worth pointing
out that the complete calculation in this setup, i.e., in general
relativity and including self-gravity and a rotating black hole, is
extremely complex and has not been done yet.

%
%

The consequences of the runaway instability could be very important
in many cases. First because this is a purely dynamical effect, so
that its time-scale is extremely short (only a few ms for a stellar
mass black hole). This means that this instability should happen
before any other processes (like the viscous transport of angular
momentum) could play a role. Secondly, because the radial mass transfer
is supposed to diverge, so that the thick disc could, in principle, be
completely destroyed.

In particular, \citet{nishida:96a} pointed out that the runaway
instability could be a severe problem for most current models of GRBs.
These models usually assume that the central engine is such a system
consisting of a black hole and a thick disc, either formed at the late
stages of the coalescence of two neutron stars 
\citep{kluzniak:98,ruffert:99,shibata:00} or after the gravitational collapse
of a massive star \citep{paczynski:86,woosley:93}. Notice that in the
latter case, the situation is somewhat more complicated because the mass of
the disc increases with time, as long as the collapse proceeds
\citep{macfadyen:99,aloy:00}. The energy which can be extracted from
this system comes from two reservoirs: the energy released by the
accretion of disc material on to the black hole and the rotational
energy of the black hole itself, which can be extracted via the
Blandford-Znajek mechanism \citep{blandford:77}. This amount of energy
(a few $10^{53}$--$10^{54}$ erg depending on the disc mass and the
black hole rotation and mass) is sufficient to power a GRB if the
energy released can be eventually converted into $\gamma$-rays with
a large efficiency of about a few percent. This number, however,
depends strongly on the central engine model and on the expected
beaming of the outflow, which is currently very poorly constrained.
Such conversion cannot be done close to the source, the luminosity
considered here being many orders of magnitude larger than the
Eddington luminosity of the system. The energy is first injected
into a very optically thick wind. This wind is accelerated via some
mechanism which is still unknown but which involves probably MHD
processes \citep{thompson:94,meszaros:97b} and becomes eventually relativistic. 
The existence of such a relativistic wind has been directly inferred from the
observations of radio scintillation in GRB 970508 \citep{waxman:98}
and is also needed to solve the so-called compactness problem and
avoid photon-photon annihilation along the line of sight. Averaged
Lorentz factors larger than 100 are required
\citep{baring:97,lithwick:01}. The observed emission is produced at
large distances from the source ($r > 10^{11}$--$10^{12}\ \mathrm{cm}$),
probably via the formation of shock waves, either within the wind itself
(internal shock model, proposed for the prompt $\gamma$-ray emission,
\citep{rees:94,kobayashi:97,daigne:98}) or caused by the deceleration of the 
wind by the external medium (external shock model, which reproduces correctly 
the afterglow emission \citep{meszaros:97a,sari:98}).

The above general scenario clearly presupposes that the black hole plus
thick disc system is stable enough to survive for a few seconds (in
particular, the internal shock model implies that the duration of the
energy release by the source has a duration comparable with the observed
duration of the GRB). However, if the runaway instability would occur,
the disc would fall into the hole in just a few milliseconds!

%
%

In order to establish the nature of instabilities in accretion flows one
must rely upon highly accurate, time-dependent, nonlinear numerical 
simulations in black hole spacetimes. These simulations are scarce 
and they have been mostly performed using a Newtonian or a 
pseudo-Newtonian potential which mimics the existence of an 
``event horizon"~\citep{paczynski:80} 
(see \citet{eggum:88,papaloizou:94,igumenshchev:96}
for simulations in the context of thick discs). Concerning relativistic 
simulations Wilson (1972) was the first to study numerically
the time-dependent accretion of matter on to a rotating black
hole. His simulations showed the formation of thick accretion discs.
In a subsequent work \citet{hawley:84a,hawley:84b} studied the
evolution and development of nonlinear instabilities in 
pressure-supported accretion discs formed as a consequence
of the spiraling infall of fluid with some amount of angular
momentum. Their constant angular momentum initial models were 
computed following the analytic theory of relativistic discs 
developed by \citet{abramowicz:78}. The code developed 
by \citet{hawley:84a,hawley:84b} was capable of keeping stable 
discs in equilibrium as well as of following the fate of initially 
unstable models. \citet{yokosawa:95} studied
the structure and dynamics of relativistic accretion discs
and the transport of energy and angular momentum in
magnetohydrodynamical accretion on to a rotating black
hole. More recently, \citet{igumenshchev:97} performed a similar
study as \citet{hawley:84a,hawley:84b}, including the
rotating black hole case and using improved numerical
methods based on Riemann solvers. They found that the structure
of the innermost part of the disc depends strongly on the
black hole spin and, at the same time, they were able to
confirm numerically the expected analytic dependence of the 
mass accretion rate with the gravitational energy gap at the 
cusp of the torus.

%
%

A time-dependent and fully relativistic study of the runaway
stability has not yet been presented in the literature. 
\citet{masuda:97} performed a time-dependent
simulation in a pseudo-Newtonian framework, using an SPH method
so that the self-gravity was taken into account. 
Our work aims at providing the first relativistic study. We have investigated the
likelihood of the instability in thick discs of constant angular momentum.
In this first investigation we present results for a Schwarzschild
black hole. The rotating case and the inclusion of accretion discs
with a distribution of angular momentum increasing outwards will be
presented elsewhere. Our simulations are performed in a fully
relativistic framework, using a 3+1 conservative formulation of
the hydrodynamic equations on a curved background \citep{banyuls:97}
and employing Godunov-type numerical methods for their solution
(see e.g. \citet{fontlr} and references therein).
As in the work of \citet{hawley:84a,hawley:84b} and \citet{igumenshchev:97}
we neglect viscous and radiative processes assuming that the flow is
isentropic. This is justified as we are only interested in phenomena
occurring on a dynamical timescale. Furthermore, as a major simplification
of the computational burden, we neglect the self-gravity of the disc.
The spacetime dynamics has hence been treated in a simple way, assuming
that the background spacetime metric is nothing but a sequence of
stationary exact black hole solutions (Schwarzschild or Kerr) of the
Einstein equations, whose dynamics is governed by the increase of the
black hole mass (and angular momentum). Such growth is controlled
only by the rate at which the mass accretes on to the hole. We are
aware that a fully consistent approach to study the problem would
imply solving the coupled system of Einstein and hydrodynamic equations
in three-dimensions. This seems still a daunting task
despite some major recent progress on numerically stable
integrations of the Einstein equations (see, e.g.,
\citet{alcubierre:00,font:01} and references therein). We believe
that, as a first step, our approach is justified and can provide some
insight on the problem.

%
%

The paper is organized as follows: Section~\ref{initial_data} reviews
briefly the theory of stationary, relativistic accretion discs of
constant angular momentum. In Section~\ref{hydro} we introduce the
equations of general relativistic hydrodynamics in the way they are
used in our code. The numerical schemes we use to solve those equations,
as well as other relevant aspects of our numerical code are described
in Section~\ref{numerics}. In Section~\ref{tests} we test the capability
of the code in keeping numerically the equilibrium of stationary models
in time-dependent simulations. The main results of the paper are presented
in Section~\ref{simulations} which contains the simulations of the runaway
instability. Finally, Section~\ref{conclusion} presents a summary of our
investigation.

Unless explicitly stated we are using geometrized units ($G=c=1$)
throughout the paper. Greek (Latin) indices run from 0 to 3 (1 to 3).
We also use the signature -+++. Usual cgs units are obtained by using
the gravitational radius of the black hole, $r_\mathrm{g}=G M_\mathrm{BH}
/c^{2}$, as unit of length. Although this paper is only concerned with the
Schwarzschild case we present all expressions for the most general setup
of a Kerr black hole in order to refer to them in the forthcoming papers
of this series. The code is already written for the Kerr metric and the
results here presented have been obtained by just setting to zero the
black hole angular momentum parameter.

\section{Stationary models of thick accretion discs}
\label{initial_data}


The theory of stationary, relativistic thick discs (or tori) of constant 
angular momentum was first derived by \citet{fishbone:76} for isentropic
discs and by \citet{abramowicz:78} for discs obeying a
barotropic equation of state (EoS). Although this
theory is by now well-established and has been presented in great detail
elsewhere, we have chosen to include here the relevant aspects and
definitions in order to facilitate the self-consistency of the paper.


\subsection{Definitions}

\subsubsection{Assumptions}

In order to build up constant angular momentum configurations for a given
metric we assume a number of conditions. Firstly, the torus is
stationary and axially symmetric. We adopt the standard spherical
coordinates $(t,r,\theta,\phi)$ in which the metric coefficients
neither depend on the time coordinate $t$ (stationarity) or on the
azimuthal coordinate $\phi$ (axisymmetry). The line element of the
spacetime is given by
\begin{equation}
ds^{2} = g_{tt} dt^{2}+2 g_{t\phi}dtd\phi+g_{\phi\phi}d\phi^{2}
+g_{rr}dr^{2}+g_{\theta\theta}d\theta^{2}\ .
\end{equation}
We adopt the following convention: the angular momentum of the black hole
is positive and the matter of the disc rotates in the positive (negative)
direction of $\phi$ for a prograde (retrograde) disc. We define the following
quantity which is a relativistic generalization of the ``distance to the
axis'':
\begin{equation}
\varpi^{2} = g_{t\phi}^{2}-g_{tt}g_{\phi\phi},
\end{equation}
which in the Newtonian limit is simply $\varpi = r\sin\theta$.

The EoS of the fluid is assumed to be barotropic,
so that $p=p(e)$ where $p$ is the pressure and $e$ is the total energy
density. We define $\rho$ as the rest mass density, $w=e+P$ as the
enthalpy and $h=w/\rho$ as the specific enthalpy. The four velocity
of the fluid is $u^{\mu}=(u^{t},0,0,u^{\phi})$ with the normalization
condition $u^{2}=u_{\mu}u^{\mu}=-1$.

\subsubsection{Von Zeipel's cylinders}

The Lagrangian angular momentum (angular momentum per unit inertial mass)
and the angular velocity are respectively defined by
\begin{equation}
l = -\frac{u_{\phi}}{u_{t}}
\end{equation}
and 
\begin{equation}
\Omega = \frac{u^{\phi}}{u^{t}}\ .
\end{equation}
From $u_{\mu} = g_{\mu\nu} u^{\nu}$ it follows that
\begin{equation}
l = -\frac{g_{t\phi}+g_{\phi\phi}\Omega}{g_{tt}+g_{t\phi}\Omega}\
\mathrm{and}\
\Omega = -\frac{g_{t\phi}+g_{tt}l}{g_{\phi\phi}+g_{t\phi}l}\ .
\label{eq:Omegal}
\end{equation}
For a barotropic EoS, the equi-$l$ and equi-$\Omega$ surfaces coincide
and they are called the von Zeipel's cylinders. Their cylinder-like
topology has been proved by \citet{abramowicz:74}. If the metric is known,
Eq.~(\ref{eq:Omegal}) allows to construct the von Zeipel's cylinder
defined by $l_{0}$ and $\Omega_{0}$ by solving the following equation
\begin{equation}
g_{tt} l_{0} + g_{t\phi}\left(1+\Omega_{0}l_{0}\right)+
g_{\phi\phi}\Omega_{0} = 0\ .
\label{eq:VonZeipel}
\end{equation}
In particular, if the distribution of the angular momentum
$l=l_\mathrm{eq}(r)$ is given in the equatorial plane $\theta = \pi/2$,
then the corresponding distribution of the angular velocity in the
equatorial plane is given by
\begin{equation}
\Omega = \Omega_\mathrm{eq}(r) =
-\frac{g_{t\phi}(r,\pi/2)+g_{tt}(r,\pi/2)l(r)}
{g_{\phi\phi}(r,\pi/2)+g_{t\phi}(r,\pi/2)l(r)}\ .
\end{equation}
The equation of the von Zeipel's cylinder intersecting the equatorial
plane at a given radial point $r=r_{0}$ is given by
Eq.~(\ref{eq:VonZeipel}) so that
\begin{eqnarray}
l^{2}(r_{0})\left[g_{tt}(r,\theta)g_{t\phi}(r_{0},\pi/2)-
g_{t\phi}(r,\theta)g_{tt}(r_{0},\pi/2)\right] & & \nonumber\\
+l(r_{0})\left[g_{tt}(r,\theta)g_{\phi\phi}(r_{0},\pi/2)-
g_{\phi\phi}(r,\theta)g_{tt}(r_{0},\pi/2)\right] & & \nonumber\\
+\left[g_{t\phi}(r,\theta)g_{\phi\phi}(r_{0},\pi/2)-
g_{\phi\phi}(r,\theta)g_{t\phi}(r_{0},\pi/2)\right]
& = & 0\ .\nonumber\\
\end{eqnarray}
Notice that in the case of the Schwarzschild metric, $g_{t\phi}=0$, and
this equation reduces to
\begin{equation}
g_{tt}(r,\theta)g_{\phi\phi}(r_{0},\pi/2)-g_{\phi\phi}(r,\theta)g_{tt}(r_{0},\pi/2) = 0
\label{eq:CylindersSchwarz}
\end{equation}
which is independent of the distribution of angular momentum.

\subsubsection{Equipotentials}

The dynamics of the gas flow is governed by the relativistic Euler
equation, whose integral form is
\begin{equation}
W-W_\mathrm{in} = \ln{(-u_{t})}-\ln{(-\left.u_{t}\right._\mathrm{in})}-
\int_{l_\mathrm{in}}^{l}\frac{\Omega dl}{1-\Omega l}\ .
\label{eq:Equilibrium}
\end{equation}
The subscript ``$\mathrm{in}$'' refers to the inner edge (in the
equatorial plane) of the disc, where the pressure vanishes. The quantity
$W$ is defined by
\begin{equation}
W-W_\mathrm{in} = -\int_{0}^{p}\frac{dp}{w}\ .
\label{eq:WEoS}
\end{equation}
In the Newtonian limit, the quantity $W$ is the total (centrifugal plus
gravitational) potential and Eq.~(\ref{eq:WEoS}) is the integral form
of the equation of hydrostatic equilibrium. If the spacetime metric is
known and if the distribution of angular momentum in the equatorial plane
is given, so that the von Zeipel's cylinders have been computed, providing
us with the value of $l$ and $\Omega$ at any given point within the
disc, then the equipotentials surfaces $W(r,\theta)$ are easily computed
from Eq.~(\ref{eq:Equilibrium}) taking into account that $u_{t}$ can be
expressed (from the 4-velocity normalization condition $u^{2}=-1$) as a
function of $l$ by
\begin{equation}
-u_{t} = \sqrt{\frac{\varpi^{2}}{g_{tt}l^{2}+2g_{t\phi}l+g_{\phi\phi}}}\ .
\label{eq:Velocity}
\end{equation}

\subsubsection{Construction of a thick disc}

In order to build a system consisting of a black hole surrounded by a thick
disc we need several parameters: the mass $M$ and the specific angular
momentum $a=J/M$ of the black hole, the distribution of the angular
momentum of the disc in the equatorial plane $l_\mathrm{eq}(r)$, the inner
radius of the disc $r_\mathrm{in}$ and the EoS of the fluid material of
the disc. The procedure is then the following:
\begin{enumerate}
\item Compute the metric coefficients in a Kerr background with free
parameters $a$ and $M$ (these coefficients are given in the next Section).
\item From the distribution of the angular momentum in the equatorial plane,
solve Eq.~(\ref{eq:VonZeipel}) to have the distribution of angular momentum
$l(r,\theta)$. Then the corresponding distribution $u_\mathrm{t}(r,\theta)$
can be evaluated from Eq.~(\ref{eq:Velocity}).
\item From the value of the inner radius $r_\mathrm{in}$, compute the
corresponding value of the angular momentum $l_\mathrm{in}=l_\mathrm{eq}
(r_\mathrm{in})$. Then the function 
\begin{equation}
F(r,\theta) = \int_{l_\mathrm{in}}^{l}\frac{\Omega dl}{1-\Omega l}
\end{equation}
can be evaluated.
\item Compute the potential 
$$
W(r,\theta)-W_\mathrm{in}=\ln{\left(-u_{t}(r,\theta)\right)}-
\ln{\left(-\left.u_{t}\right._\mathrm{in}\right)}-F(r,\theta)\ .
$$ 
The constant $W_\mathrm{in}$ is fixed by the convention that
$\lim_{r\to+\infty} W(r,\theta)=0$.
\item Compute all hydrodynamical quantities ($p$, $\rho$, $e$, $w$, etc)
from the EoS and Eq.~(\ref{eq:WEoS}). In the following , we will only
consider the particular case of isentropic fluids where 
\begin{equation}
\int_{0}^{p}\frac{dp}{w} = \ln{\frac{h}{h_\mathrm{in}}}\ .
\label{eq:FillMatter}
\end{equation}
\end{enumerate}
If the self-gravity of the disc is neglected, the procedure stops here.
Otherwise, from the distribution of matter-energy we have computed, we
need to solve the Einstein field equations to evaluate the new metric
coefficients $g_{\mu\nu}$. Then the procedure starts again at step 2.
Such cycles are repeated until convergence.

Once a disc has been built, one can estimate its mass with the expression
\begin{equation}
m = \int\!\!\!\!\int\!\!\!\!\int\! \left(T^{\phi}_{\phi}+T^{r}_{r}+
T^{\theta}_{\theta}-T^{t}_{t}\right)\sqrt{-g}\ dV\ ,
\end{equation}
where $T^{\mu \nu}$ is the stress-energy tensor which, in the case of
a perfect fluid is defined by $T^{\mu \nu}=\rho h u^{\mu} u^{\nu}+
p g^{\mu\nu}$. In all the models we will consider in the following,
we have $p \ll e$ so that the mass of the disc can be accurately
approximated by the rest-mass
\begin{eqnarray}
m & \simeq & \int\!\!\!\!\int\!\!\!\!\int\!\! 
\rho\left(u_{\phi}u^{\phi}-u_{t}u^{t}\right)\sqrt{-g}\ dV\nonumber\\
& \simeq & \int\!\!\!\!\int\!\!\!\!\int\!\! 
\frac{1+\Omega l}{1-\Omega l} \rho \sqrt{-g}\ dV\ .
\end{eqnarray}

\begin{figure}
\centerline{\psfig{file=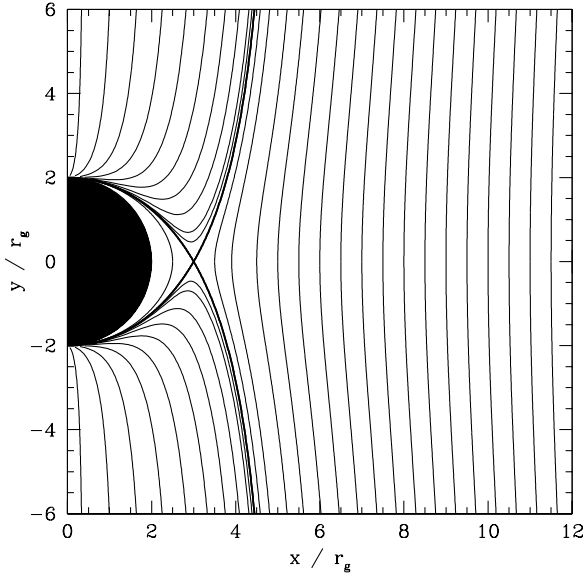,width=8cm}}
\caption{The von Zeipel's cylinders for the Schwarzschild metric. 
The critical cylinder (thick line) has a cusp located at $r=3$ in
the equatorial plane.}
\label{fig:VonZeipel}
\end{figure}

\subsection{Constant angular momentum thick discs orbiting a
Schwarzschild black hole}

In this subsection we describe in detail the particular case we have
considered for the time-dependent simulations of this paper: the black
hole is non-rotating (a Schwarzschild black hole), the self-gravity
of the disc is neglected, the angular momentum $l$ is constant
and the disc obeys a polytropic EoS with
\begin{equation}
P = \kappa \rho^{\gamma},
\label{eq:Polytrop}
\end{equation}
with $\kappa$ being the polytropic constant and $\gamma$ the adiabatic
exponent. For commodity, we further assume that the mass of the black
hole is equal to unity. The metric coefficients are given by
\begin{eqnarray}
g_{tt} & = & -\left(1-\frac{2}{r}\right),\\
g_{t\phi} & = & 0,\\
g_{\phi\phi} & = & r^{2}\sin^{2}\theta,
\end{eqnarray}
and the distance to the axis is simply $\varpi^{2}=r(r-2)\sin^{2}\theta$.
As the black hole is non-rotating, we also consider only positive values
of the angular momentum (prograde discs).

The von Zeipel's cylinders are independent of $l$. Using
Eq.~(\ref{eq:CylindersSchwarz}), the cylinder intersecting the equatorial
plane at $r=r_{0}$ is given by
\begin{equation}
(r_{0}-2) r^{3}\sin^{2}\theta-r_{0}^{3} (r-2) = 0\ .
\end{equation}
The result is shown in Fig.~\ref{fig:VonZeipel}. Notice the cusp located
at $r=3$ in the equatorial plane (thick line).

Since the angular momentum $l$ is constant, the function $F(r,\theta)$
vanishes. Using Eq.~(\ref{eq:Velocity}), the component $u_{t}$ of the
4-velocity is given by
\begin{equation}
-u_{t} = r\sin{\theta} \sqrt{\frac{r-2}{r^{3}\sin^{2}\theta-(r-2)l^{2}}}
\end{equation}
and the potential $W$ reads
\begin{equation}
W(r,\theta) = \frac{1}{2}\ln{\frac{r^{2}(r-2)\sin^{2}\theta}
{r^{3}\sin^{2}\theta-l^{2}(r-2)}},
\end{equation}
where we have used the condition $W(r,\theta)\to 0$ for $r\to+\infty$
to eliminate $r_\mathrm{in}$.

We describe now the equipotentials which are either closed ($W<0$) or
open ($W>0$). The marginal case $W=0$ is closed at infinity. The geometry
of the equipotentials is fixed by the value of $l$. We impose that $W(r)$
is defined everywhere outside the horizon in the equatorial plane, i.e.
that the term $r^{3}-l^{2}(r-2)$ never vanishes. Then
$l<l_\mathrm{max}=3\sqrt{3}\simeq 5.20$.


\begin{table*}
\caption{The possible equipotential configurations for a Schwarzschild
black hole and a constant angular momentum disc.}
\begin{tabular}{llllll}
\hline
Angular momentum & $r_\mathrm{cusp}$ & $W_\mathrm{cusp}$ & $r_\mathrm{center}$ & 
$W_\mathrm{center}$ & Comments\\
\hline
\hline
$l<l_\mathrm{ms}\simeq 3.67$ & & & & & No cusp. No center. Disc infinite.\\
$l=l_\mathrm{ms}\simeq 3.67$ & $r_\mathrm{cusp}=r_\mathrm{ms}=6$ & $<0$ & 
$r_\mathrm{center}=r_\mathrm{ms}=6$ & $<0$ & Cusp = center. Disc infinite.\\
$l_\mathrm{ms} < l < l_\mathrm{mb}$ & $r_\mathrm{mb} < r_\mathrm{cusp} < r_\mathrm{ms} 
$ & $<0$ & $r_\mathrm{center} > r_\mathrm{ms}$ & $<0$ & Cusp. Center. Disc closed.\\
$l=l_\mathrm{mb}=4$ & $r_\mathrm{cusp}=r_\mathrm{mb}=4$ & $0$ & $r_\mathrm{center}\simeq 
10.47$ & $<0$ & Cusp. Center. Disc closed at infinity.\\
$l_\mathrm{mb} < l < l_\mathrm{max}$ & $r_\mathrm{cusp}<r_\mathrm{mb}$ & $>0$ & 
$r_\mathrm{center}\ga 10.47$ & $<0$ & Cusp. Center. Disc infinite$^{*}$.\\
$l = l_\mathrm{max}\simeq 5.20$ & $r_\mathrm{cusp}=3$ & $+\infty$ & 
$r_\mathrm{center}\simeq 22.39$ & $<0$ & Cusp marginally defined. Center. Disc infinite$^{*}$.\\ 
\hline
\end{tabular}
\begin{flushleft}
$^{*}$\,Some closed equipotentials are still present around the center.
\end{flushleft}
\label{tab:Equipotentials}
\end{table*}

It can be easily shown that the presence of a cusp in the equatorial
plane is related to the solutions of 
\begin{equation}
l_\mathrm{K}(r) = l\ ,
\label{eq:CuspCenter}
\end{equation}
where $l_\mathrm{K}(r)$ is the Keplerian angular momentum of a particle
located at a radius $r$ in the equatorial plane. It is given by
\begin{equation}
l_\mathrm{K}(r) = \frac{r\sqrt{r}}{r-2}\ .
\end{equation}
Two useful radii can be defined, the radius of the last (marginally) stable 
orbit $r_\mathrm{ms}=6$ (corresponding to the minimum of $l_\mathrm{K}(r)$),
and the radius of the last (marginally) bound orbit $r_\mathrm{mb}=4$.
The corresponding values of $l_\mathrm{K}$ are $l_\mathrm{ms} =
\frac{3\sqrt{6}}{2}\simeq 3.67$ and $l_\mathrm{mb} = 4$. The properties
of $W(r,\theta)$ are given in Table \ref{tab:Equipotentials} for the
different possible values of $l$. The corresponding equipotentials are
drawn in Fig.~\ref{fig:Equipotentials} (right panel). The angular momentum
and potential in the equatorial plane are also drawn on the left panel
of Fig.~\ref{fig:Equipotentials}. It is clear that the most interesting
case is case (3) where $l_\mathrm{ms} < l < l_\mathrm{mb}$ so that there
exist a cusp, a center, and the equipotential of the cusp is closed.
From Eq.~(\ref{eq:Polytrop}) and Eq.~(\ref{eq:FillMatter}) we have indeed 
\begin{equation}
W-W_\mathrm{in} = -\ln{h} = -\ln{\left(1+\frac{\gamma}
{\gamma-1}\kappa\rho^{\gamma-1}\right)},
\end{equation}
so that the matter can fill only the part where $W\le W_\mathrm{in}$.
The density and the pressure are then easily determined from $h$:
\begin{eqnarray}
h & = & e^{W_\mathrm{in}-W},\\
\rho & = & \left(\frac{\gamma-1}{\gamma}\frac{e^{W_\mathrm{in}-W}-1}
{\kappa}\right)^{\frac{1}{\gamma-1}},\\
P & = & \kappa\left(\frac{\gamma-1}{\gamma}\frac{e^{W_\mathrm{in}-W}-1}
{\kappa}\right)^{\frac{\gamma}{\gamma-1}}.
\end{eqnarray}
It is then possible to adjust the value of $l$ and $r_\mathrm{in}$ to fix
the mass of the disc, which is given by 
\begin{equation}
m \simeq 2\pi \int\!\!\!\!\int\!\! 
\rho\frac{r^{3}\sin^{2}\theta+(r-2)l^{2}}{r^{3}\sin^{2}\theta-(r-2)l^{2}} 
r^{2}\sin\theta d\theta dr\ .
\end{equation}
If one imposes the condition that the disc is exactly filling its Roche
lobe, the inner radius $r_\mathrm{in}=r_\mathrm{cusp}$ is fixed. Otherwise,
instead of specifying $r_\mathrm{in}$, one could prefer another parameter,
such as the potential barrier (energy gap) at the inner edge defined as 
\begin{equation}
\Delta W_\mathrm{in} = W_\mathrm{in}-W_\mathrm{cusp}\ .
\end{equation}
The case $\Delta W_\mathrm{in} < 0 $ corresponds to a disc inside its
Roche lobe. No mass transfer is possible. The case $\Delta W_\mathrm{in} >
0$ corresponds to a disc overflowing its Roche lobe: mass transfer is
possible at the cusp. An analytic estimation for the mass flux (flux of
rest mass density) was derived by \citet{abramowicz:78}
for this last case, showing the following dependence:
\begin{equation}
\dot{m} \propto \Delta W_\mathrm{in}^{\frac{\gamma}{\gamma-1}}.
\label{eq:mdotvsdw}
\end{equation}

\begin{figure*}
\begin{tabular}{cc}
\begin{minipage}[t]{7.25cm}
\psfig{file=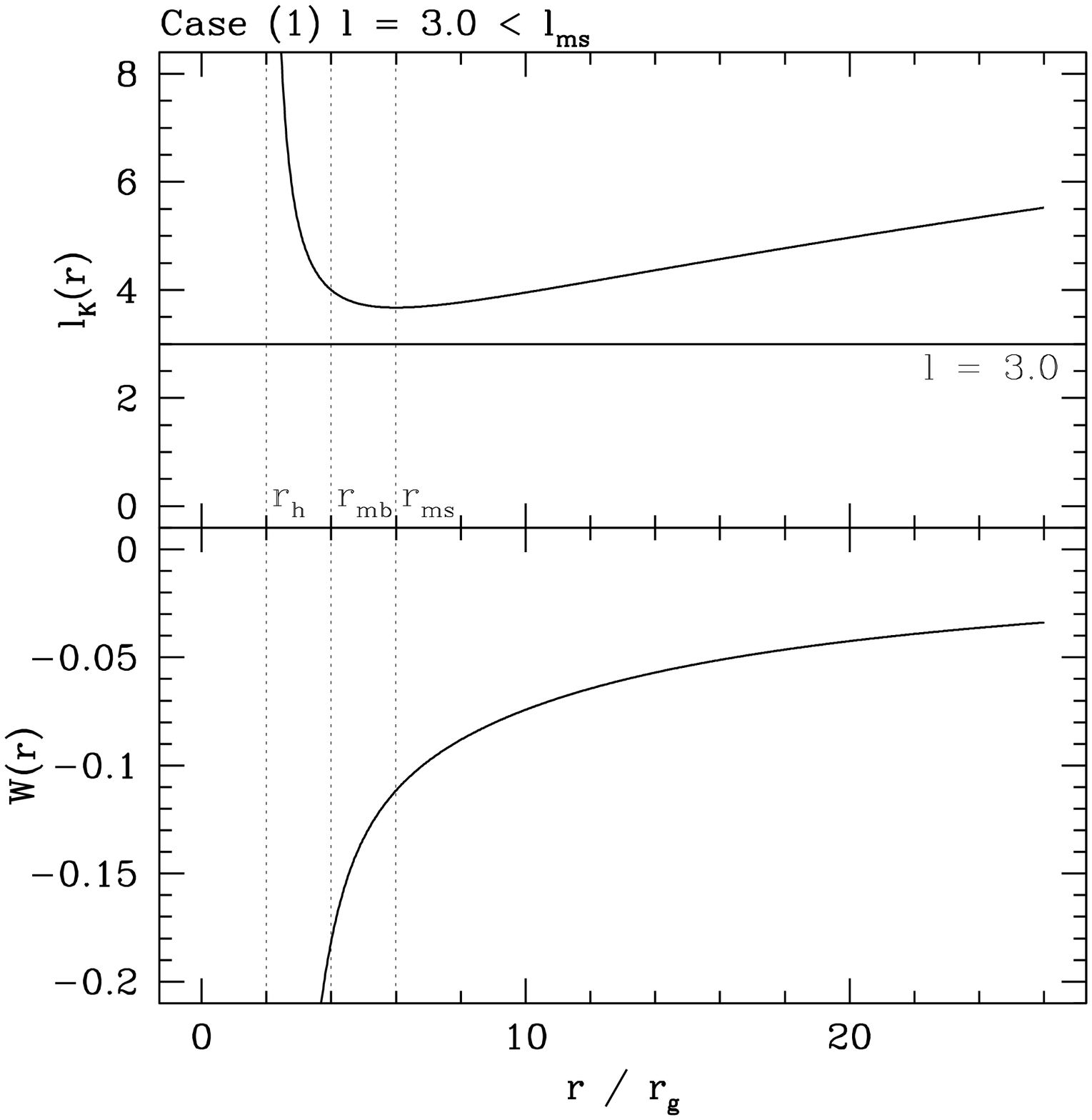,width=7.25cm}
\end{minipage} &
\begin{minipage}[t]{7.25cm}
\psfig{file=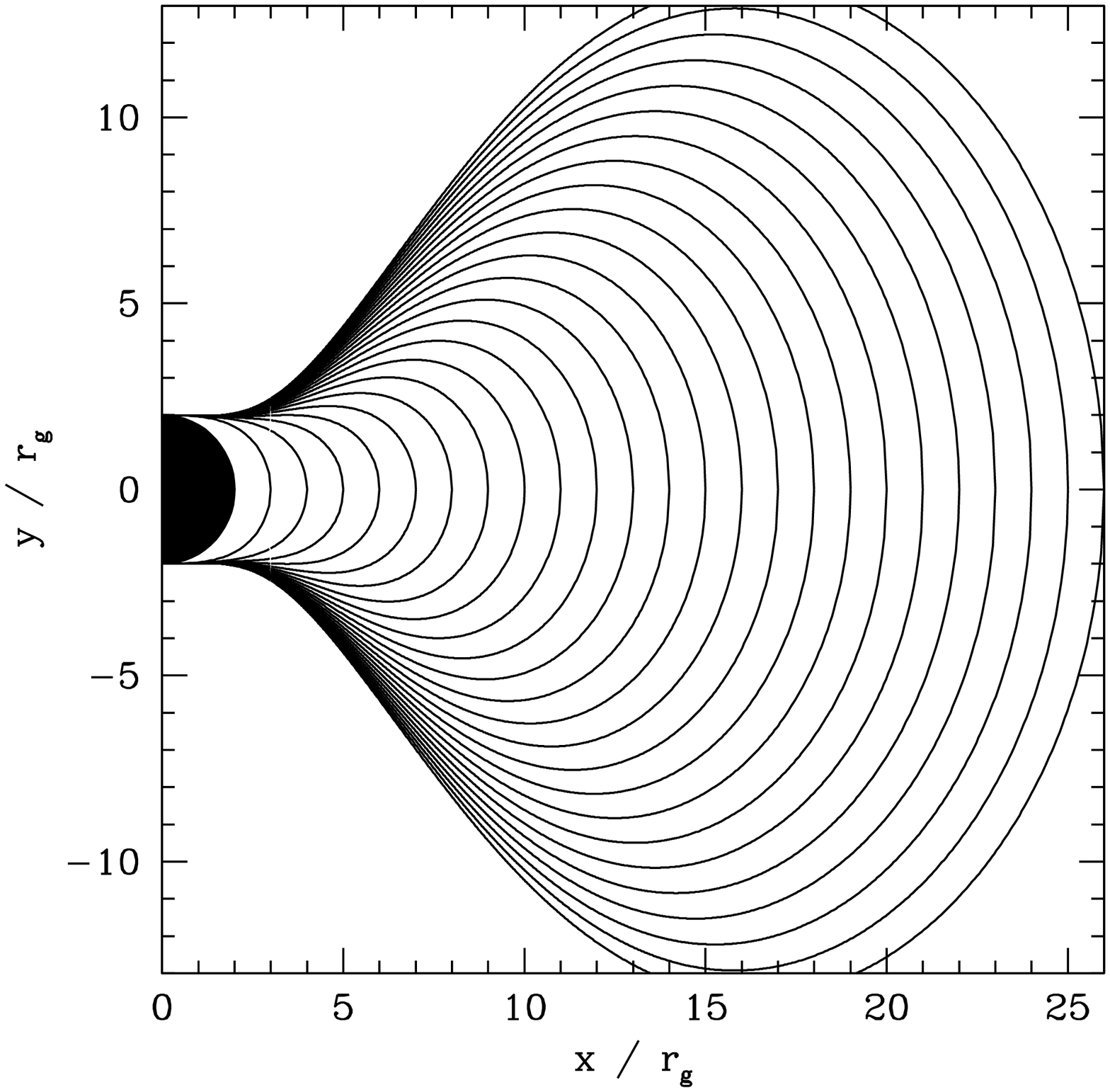,width=7.25cm}
\end{minipage}\\
\begin{minipage}[t]{7.25cm}
\psfig{file=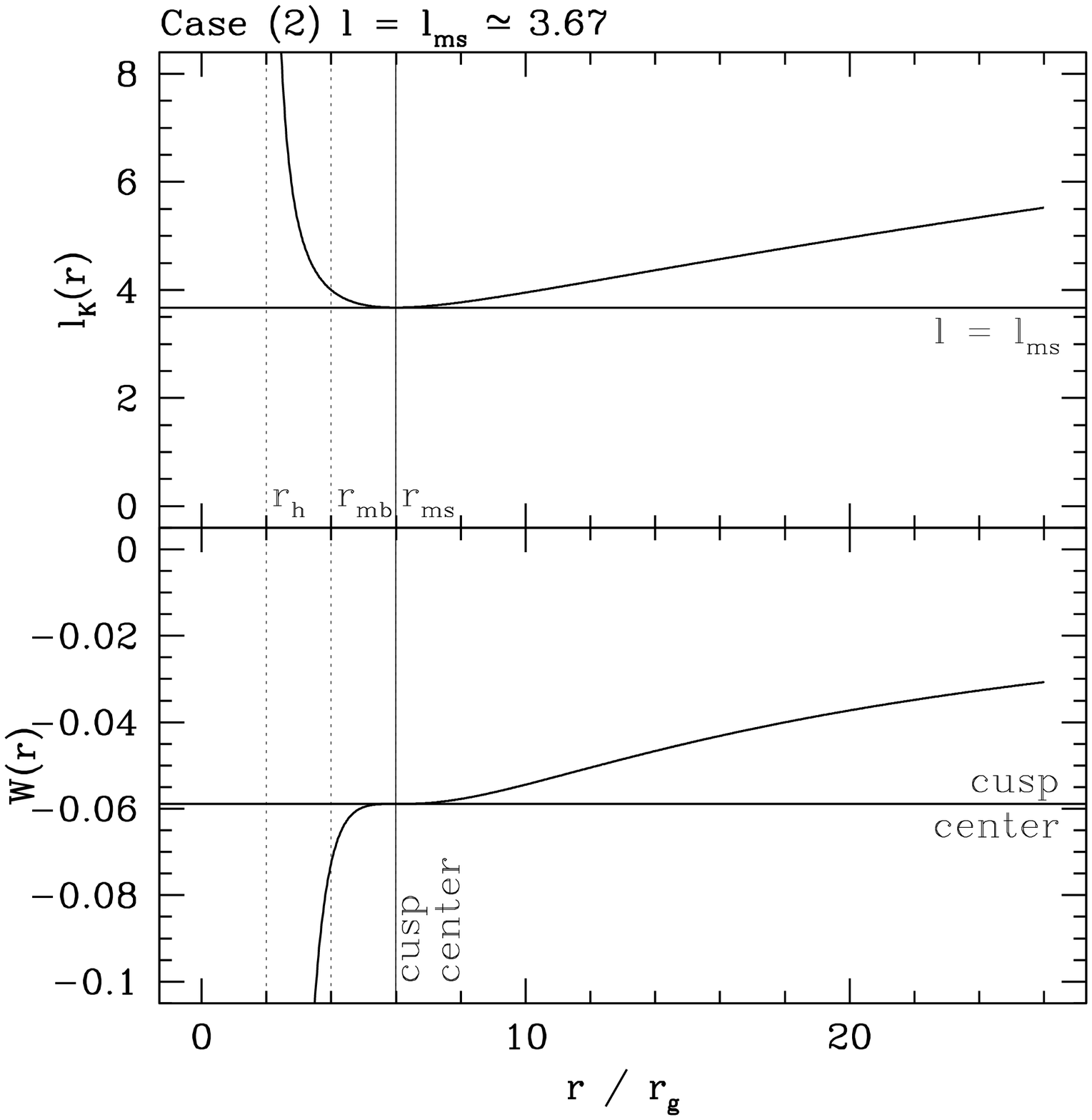,width=7.25cm}
\end{minipage} &
\begin{minipage}[t]{7.25cm}
\psfig{file=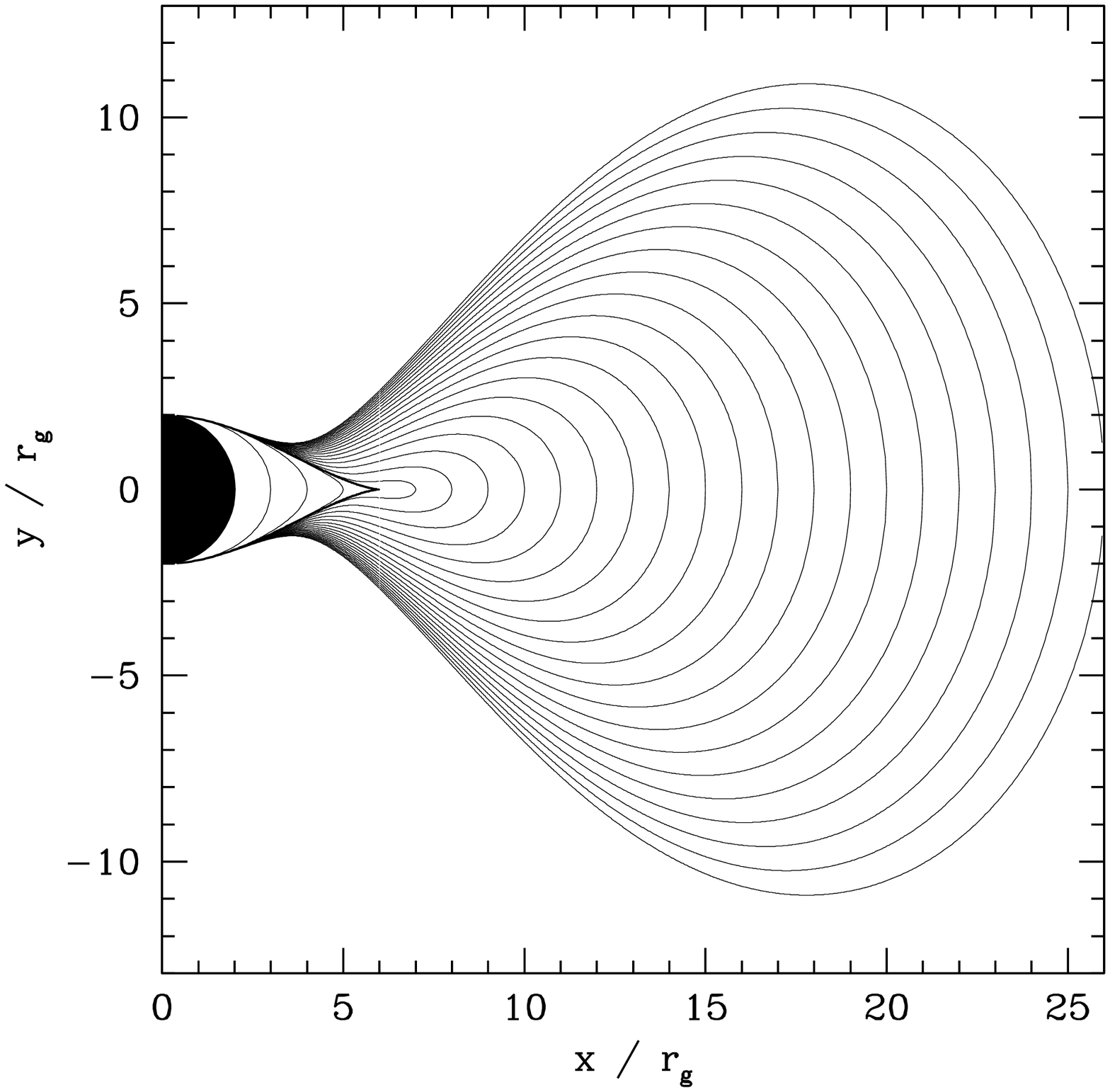,width=7.25cm}
\end{minipage}\\
\begin{minipage}[t]{7.25cm}
\psfig{file=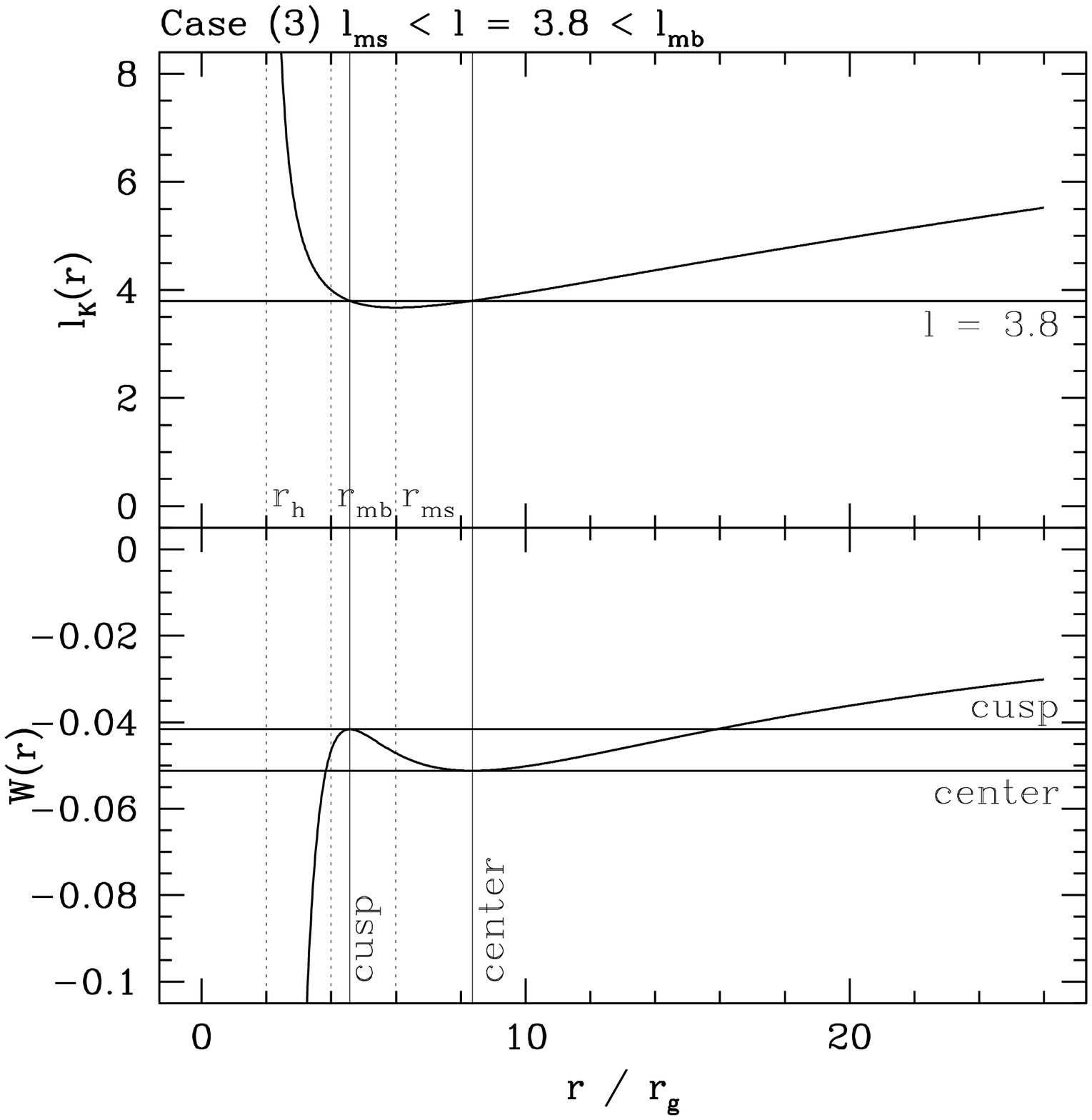,width=7.25cm}
\end{minipage} &
\begin{minipage}[t]{7.25cm}
\psfig{file=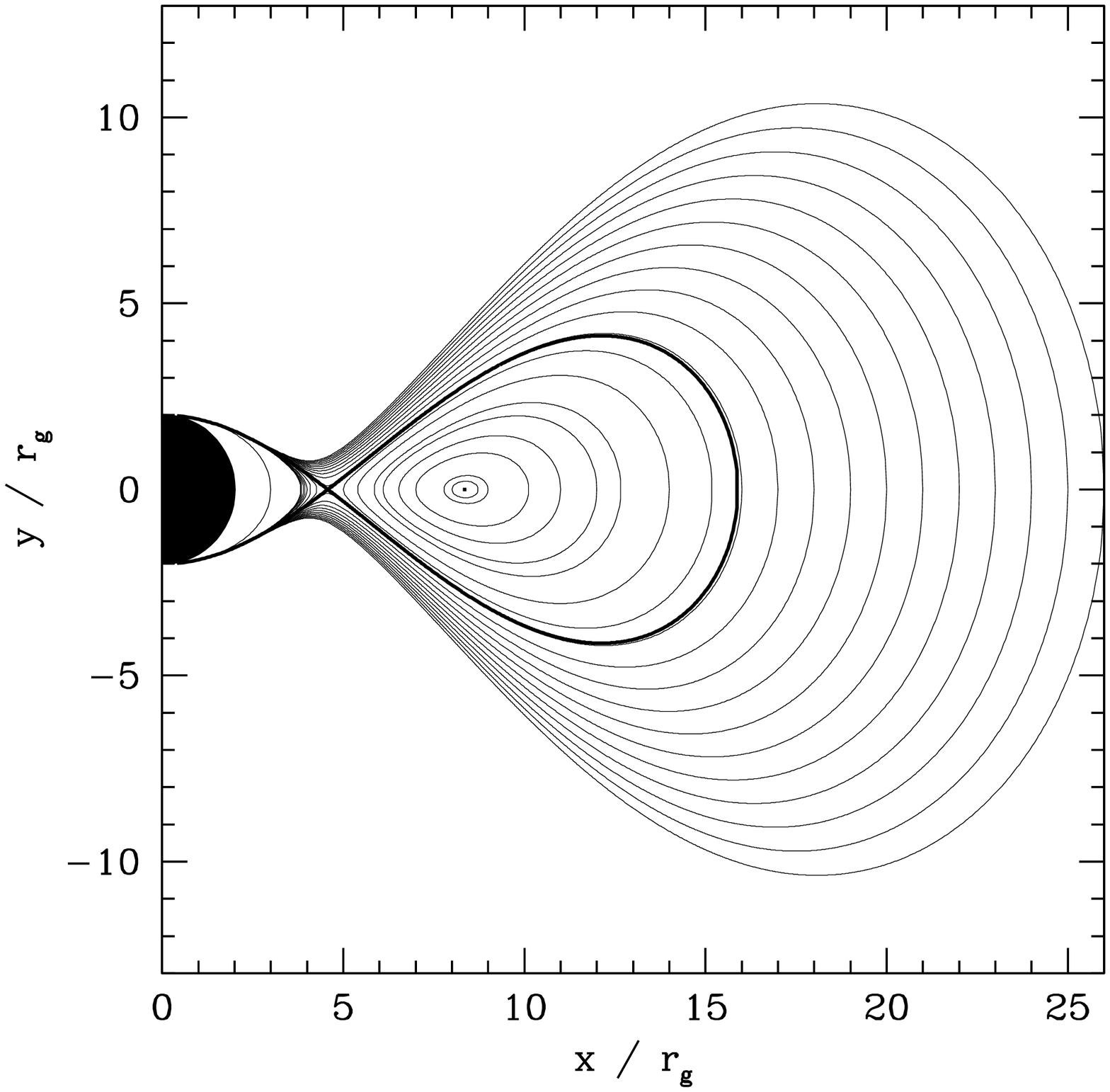,width=7.25cm}
\end{minipage}
\end{tabular}
\caption{Constant angular momentum thick disc orbiting a Schwarzschild
black hole: Equipotentials for different values of $l$ (right column). 
The thick line corresponds to the equipotential of the cusp and a thick dot marks the
center. The physically interesting case (corresponding to a thick torus)
is case (3).}
\label{fig:Equipotentials}
\end{figure*}

\begin{figure*}
\begin{tabular}{cc}
\hspace{1cm}
\begin{minipage}[t]{7.25cm}
\psfig{file=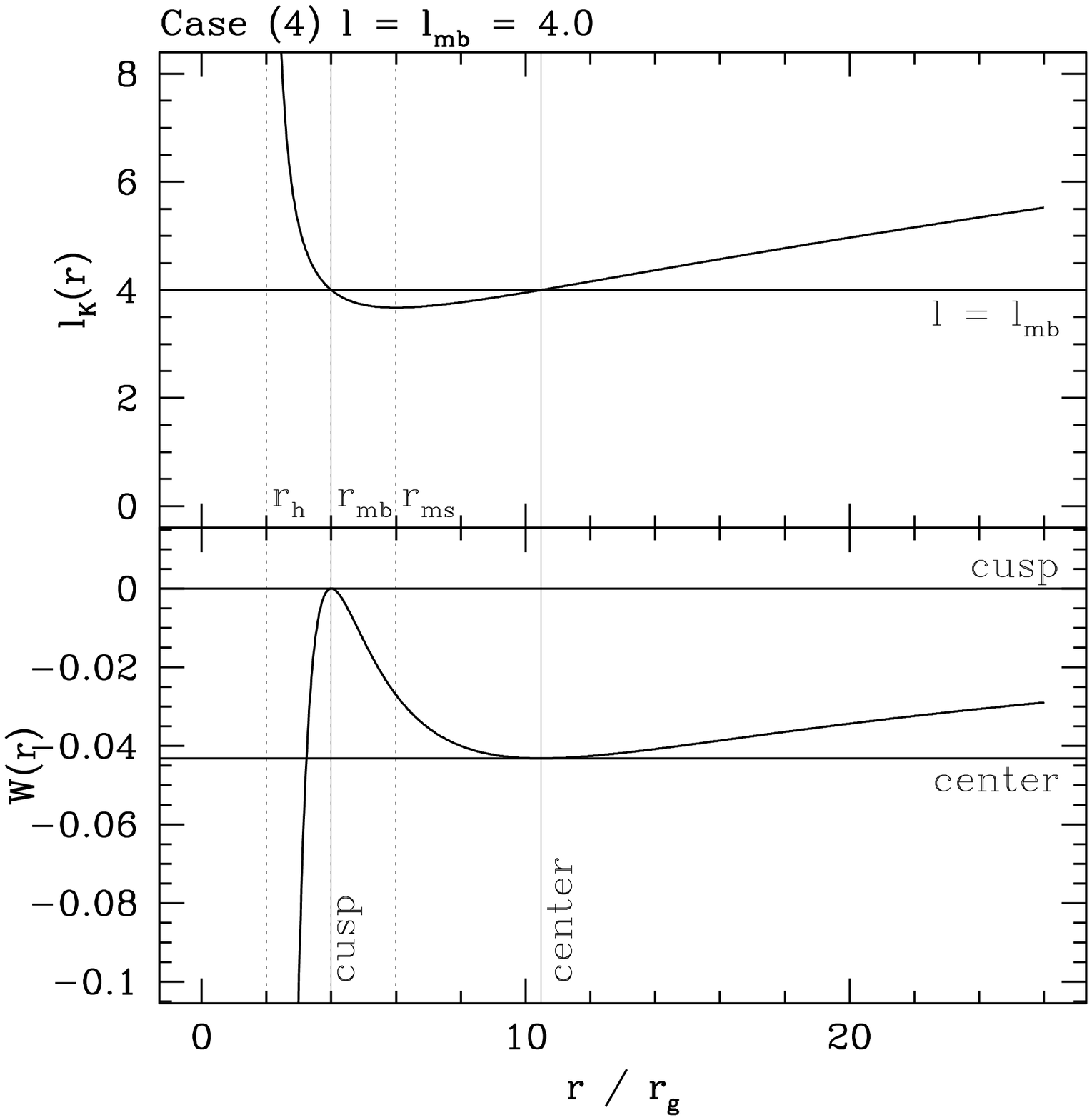,width=7.25cm}
\end{minipage} &
\begin{minipage}[t]{7.25cm}
\psfig{file=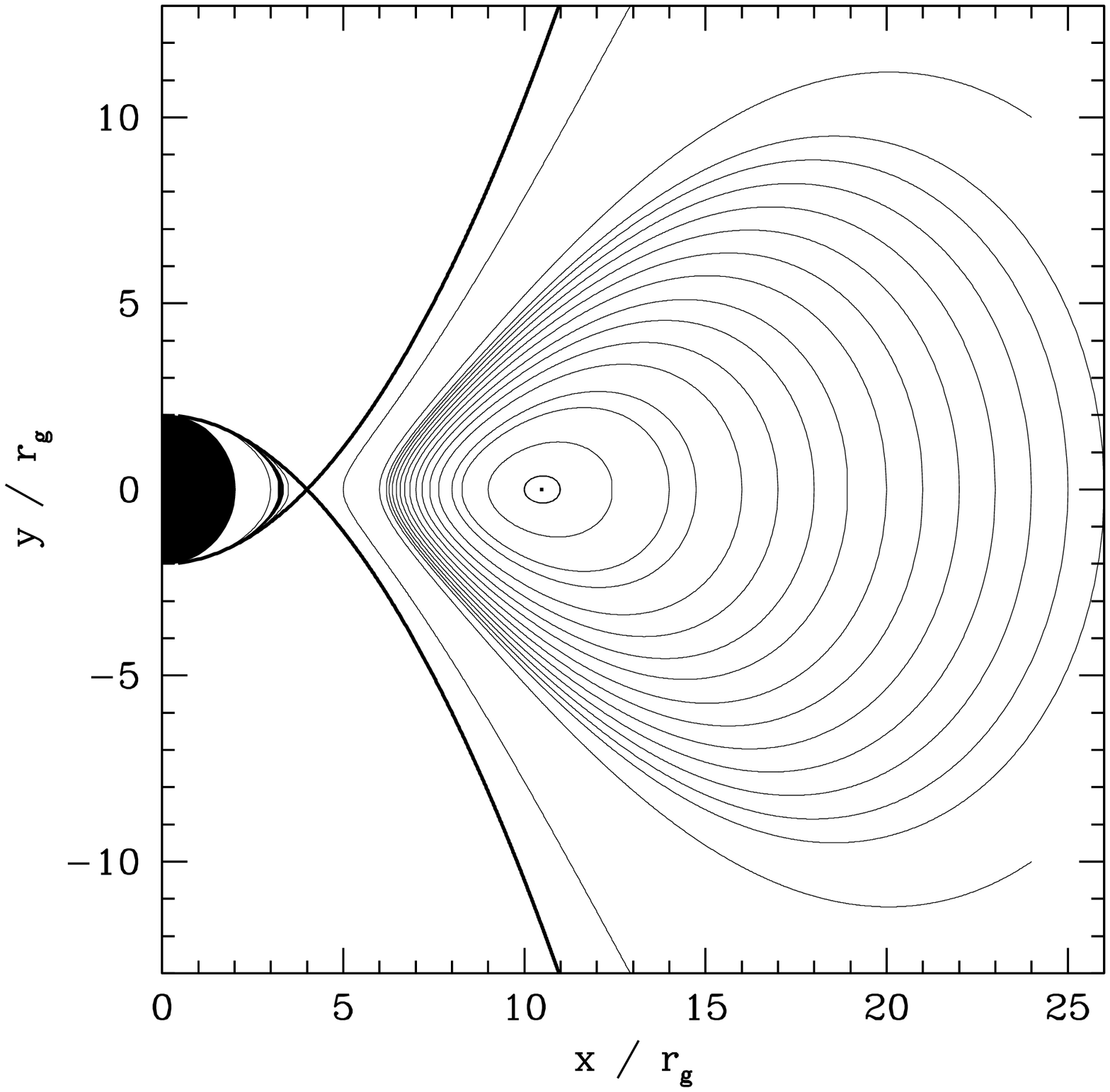,width=7.25cm}
\end{minipage}\\
\hspace{1cm}
\begin{minipage}[t]{7.25cm}
\psfig{file=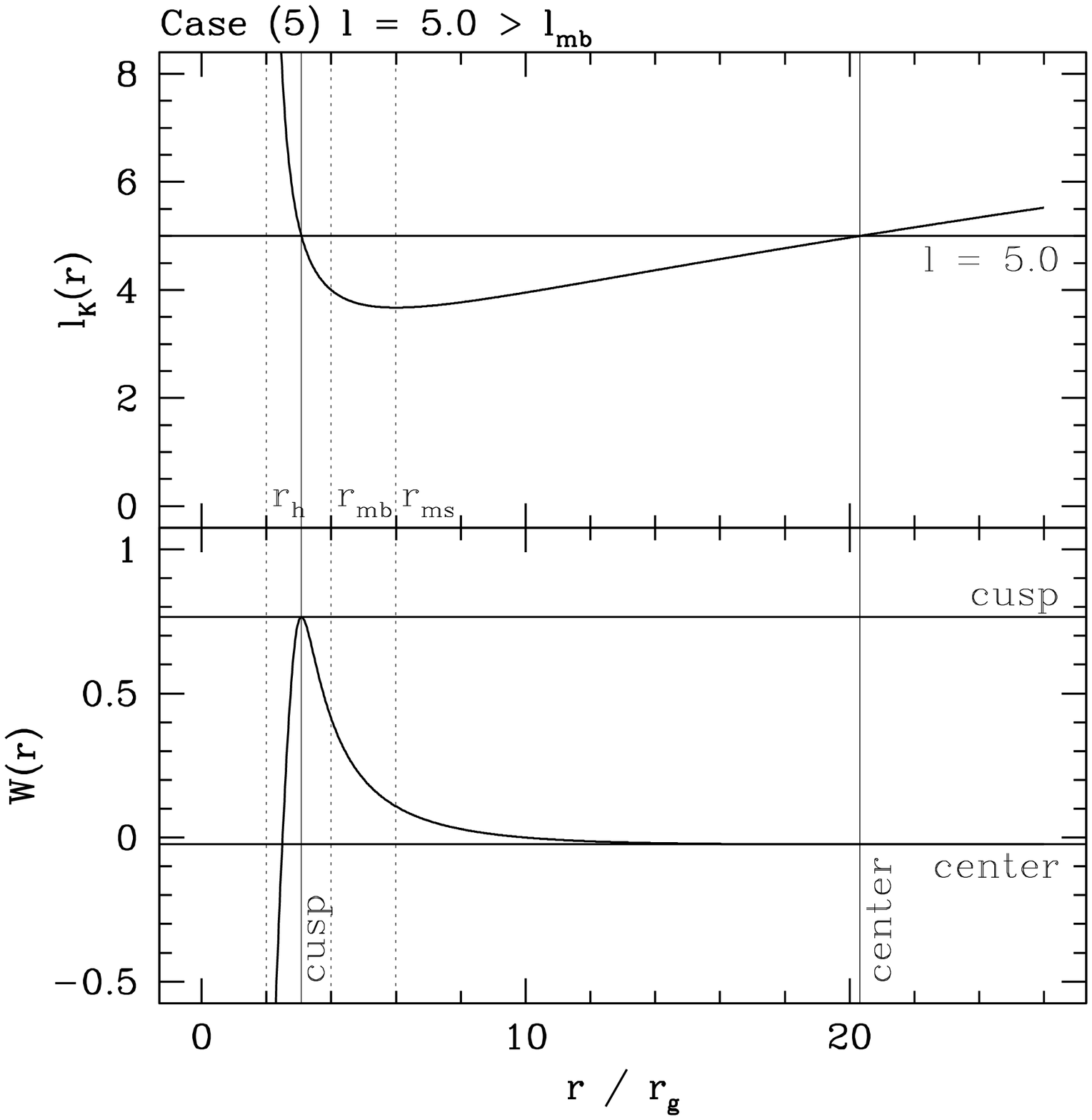,width=7.25cm}
\end{minipage} &
\begin{minipage}[t]{7.25cm}
\psfig{file=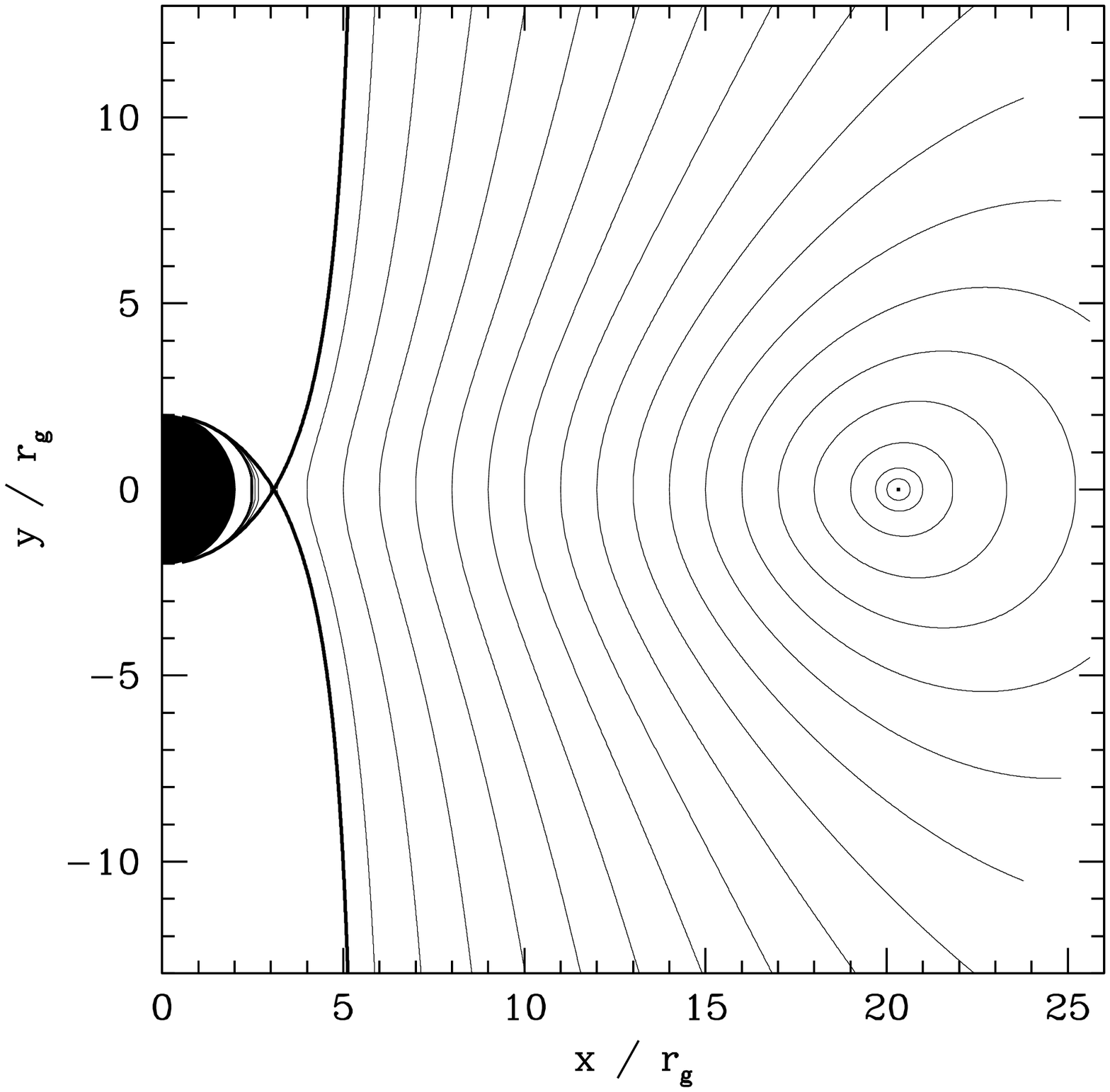,width=7.25cm}
\end{minipage}\\
\hspace{1cm}
\begin{minipage}[t]{7.25cm}
\psfig{file=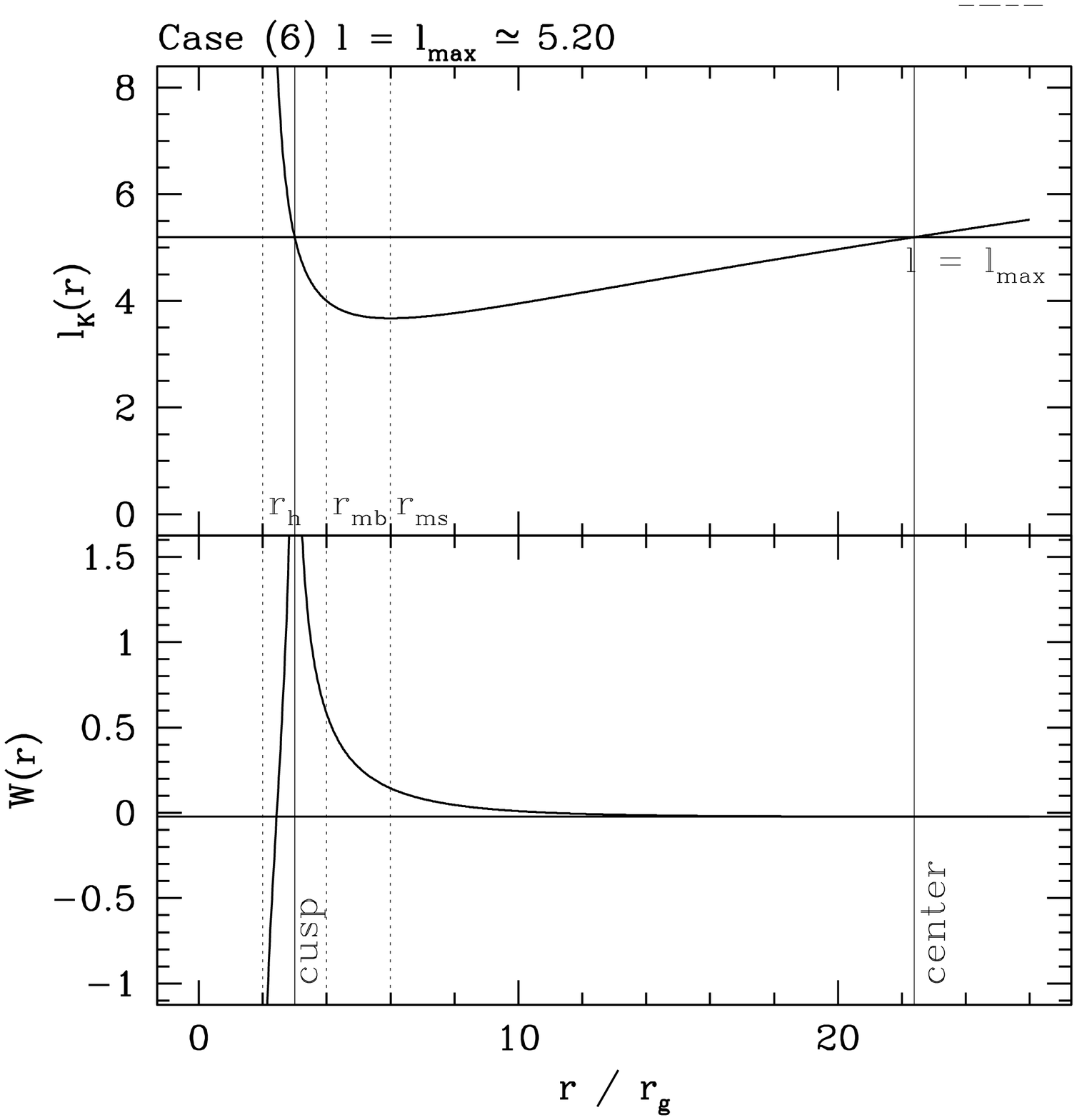,width=7.25cm}
\end{minipage} &
\begin{minipage}[t]{7.25cm}
\psfig{file=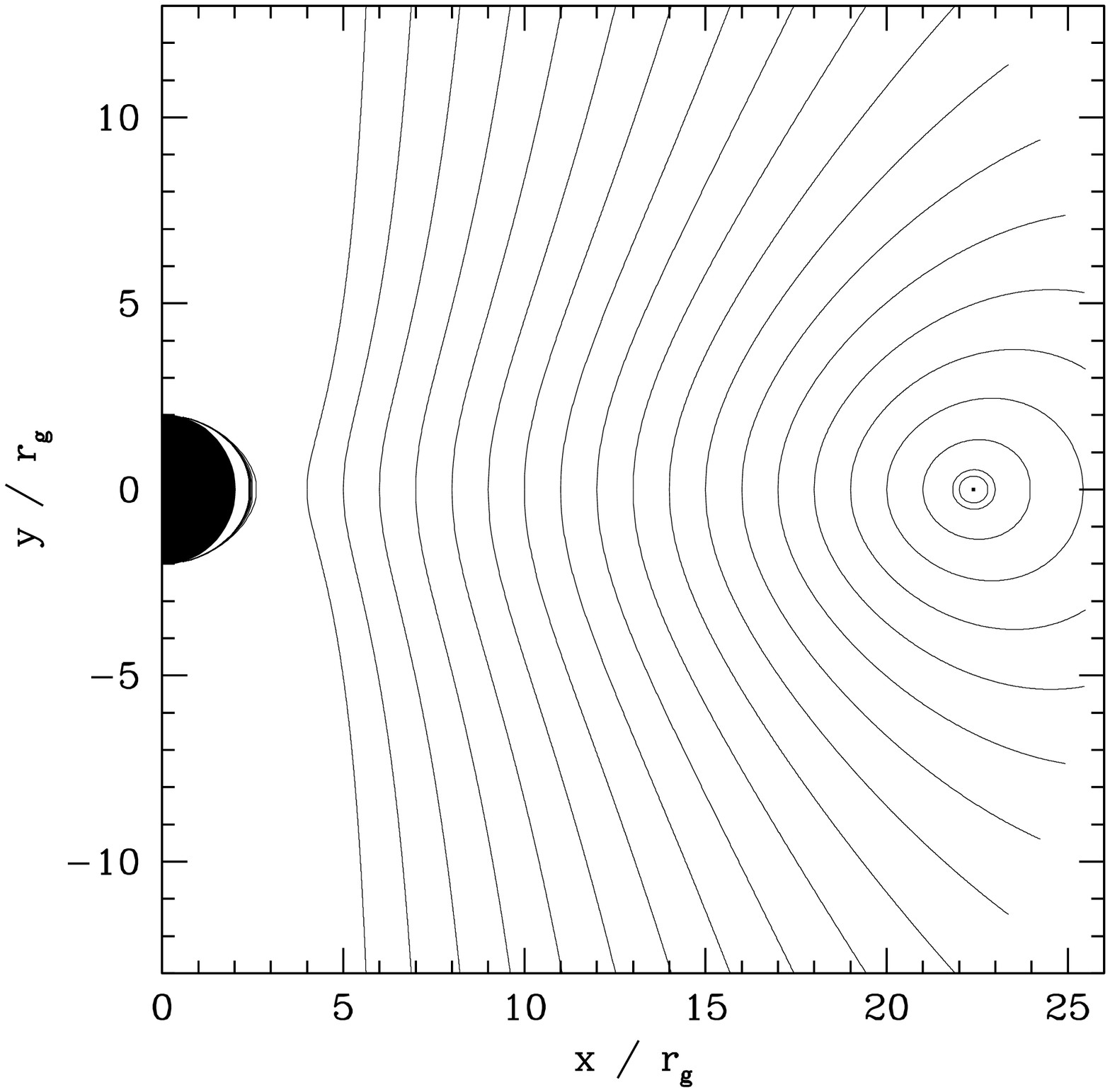,width=7.25cm}
\end{minipage}
\end{tabular}
\contcaption{Constant angular momentum thick disc orbiting a Schwarzschild
black hole: Equipotentials for different values of $l$ (right column). The 
thick line corresponds to the equipotential of the cusp and a thick dot 
marks the center.}
\end{figure*}

\section{Hydrodynamic equations in a Kerr background}
\label{hydro}

Our purpose is to evolve in time the initial data described in the
previous Section. In order to do so we present in this Section
the formulation of the general relativistic hydrodynamic equations
in the form they have been implemented in our numerical code.

Using Boyer-Lindquist $(t,r,\theta,\phi)$ coordinates, the Kerr line
element, $ds^2=g_{\mu\nu} dx^{\mu} dx^{\nu}$, reads
\begin{eqnarray}
ds^2 &=& - { {\Delta - a^2\sin^2\theta} \over {\varrho^2} } dt^2 -
       2a { {2Mr\sin^2\theta} \over {\varrho^2} } dt d\phi     
\nonumber \\ &+&
       { {\varrho^2} \over {\Delta} } dr^2                     + 
       \varrho^2 d\theta^2 +
       { {\Sigma} \over {\varrho^2} }
       \sin^2\theta d\phi^2,
\label{blform}
\end{eqnarray}
\noindent
with the definitions:
\begin{eqnarray}
\Delta &\equiv& r^2 - 2Mr + a^2,
\\
\varrho^2 &\equiv& r^2 + a^2\cos^2\theta,
\label{metrho}
\\
\Sigma &\equiv& (r^2+a^2)^2 - a^2\Delta\sin^2\theta,
\end{eqnarray}
\noindent
where $M$ is the mass of the black hole and $a$ is the black hole angular
momentum per unit mass ($J/M$). Notice that the geometrical factor $\varrho$ 
has not to be confused with the rest-mass density of the fluid, $\rho$. 
The above metric, Eq.~(\ref{blform}), describes 
the spacetime exterior to a rotating and non-charged black hole. The metric 
has a coordinate singularity at the roots of the equation
$\Delta=0$, which correspond to the horizons of a rotating black hole,
$r=r_{\pm}=M\pm(M^2-a^2)^{1/2}$. The ``distance to the rotation axis'' 
introduced in the previous Section is given by
$\varpi^{2} = g_{t\phi}^{2}-g_{tt}g_{\phi\phi} = \Delta\sin^2\theta $.

The \{3+1\} decomposition (see, e.g., \citet{misner:73})
of this form of the metric leads to a spatial 3-metric $\gamma_{ij}$ with
non-zero elements given by $\gamma_{rr} = \varrho^{2}/\Delta, 
\gamma_{\theta\theta} = \varrho^2, \gamma_{\phi\phi} = {\Sigma} /
{\varrho^2}  \sin^2\theta$. In addition,
the azimuthal {\em shift vector} $\beta_{\phi}\equiv g_{t\phi}$ is given by
\begin{equation}
\beta_{\phi} = -\frac{2aMr\sin^2\theta}{\varrho^{2}},
\end{equation}
and the {\em lapse function} is given by
\begin{equation}
\alpha = \left( \frac{\varrho^{2} \Delta}{\Sigma} \right)^{1/2}.
\end{equation}

The equations of general relativistic hydrodynamics are obtained from
the local conservation laws of density current $J^{\mu}$ and
stress-energy $T^{\mu\nu}$
\begin{eqnarray}
\nabla_{\mu}J^{\mu}&=&0, \\
\nabla_{\mu}T^{\mu \nu}&=&0,
\end{eqnarray}
\noindent
with
\begin{eqnarray}
J^{\mu}&=&\rho u^{\mu},
\\
T^{\mu \nu}&=&\rho hu^{\mu}u^{\nu} + p g^{\mu \nu},
\end{eqnarray}
\noindent
for a general EoS of the form $p=p(\rho,\varepsilon)$,
$\varepsilon$ being the specific internal energy. The specific enthalpy
is defined as $h=1+\varepsilon+{P}/{\rho}$. Furthermore, $\nabla_{\mu}$
is the covariant derivative associated with the four-dimensional metric
$g_{\mu\nu}$ and $u^{\mu}$ is the fluid 4-velocity. The above expression
of the stress-energy tensor corresponds to that of a perfect fluid.

Following the general approach laid out in \cite{banyuls:97}, after
the choice of an appropriate vector of conserved quantities, the
general relativistic hydrodynamic equations can be written as a first-order
flux-conservative hyperbolic system. In axisymmetry ($\partial_{\phi}=0$)
and with respect to the Kerr metric such system adopts the form
\begin{eqnarray}
 {{\partial {\bf U}({\bf w})} \over {\partial t}} +
 {{\partial (\alpha {\bf F}^{r}({\bf w}))} \over {\partial r}} +
 {{\partial (\alpha {\bf F}^{\theta}({\bf w}))} \over {\partial \theta}}
 = {\bf S}({\bf w}).
\label{system}
\end{eqnarray}
\noindent
In this equation the vector of (physical) {\it primitive variables} is
defined as
\begin{eqnarray}
{\bf w} = (\rho, v_{r}, v_{\theta}, v_{\phi}, \varepsilon),
\end{eqnarray}
\noindent
where $v_i (i=r,\theta,\phi)$ is the fluid 3-velocity, defined as
$v^i=u^i/\alpha u^t + \beta^i/\alpha$, with $v_i=\gamma_{ij}v^j$.
On the other hand, the state vector (evolved quantities) in 
Eq.~(\ref{system}) is
\begin{eqnarray}
{\bf U}({\bf w})  =   (D, S_r, S_{\theta}, S_{\phi}, \tau).
\end{eqnarray}
\noindent
The explicit relations between the two sets of variables, ${\bf U}$ and
${\bf w}$, are
\begin{eqnarray}
D &=&  \rho \Gamma,
\nonumber
\\
S_j &=&  \rho h \Gamma^2 v_j \,\, (j=r, \theta,\phi),
\label{elf}
\\
\tau &=&  \rho h \Gamma^2 - p - D,
\nonumber
\label{eq:DefLabFrame}
\end{eqnarray}
\noindent
with $\Gamma$ being the Lorentz factor, $\Gamma\equiv\alpha u^t =
(1-v^2)^{-1/2}$, with $v^2=\gamma_{ij}v^i v^j$.  The specific form of the
fluxes, ${\bf F}^{i}$, and the source terms, {\bf S}, read
\begin{eqnarray}
{\bf F}^{r}({\bf w})  =   \left(D v^{r},
 S_r v^{r} + p, S_{\theta} v^{r}, S_{\phi} v^{r},
(\tau + p) v^{r} \right),
\end{eqnarray}
\noindent
\begin{eqnarray}
{\bf F}^{\theta}({\bf w})  =   \left(D v^{\theta}, 
S_{r} v^{\theta}, 
S_{\theta} v^{\theta} + p,
S_{\phi} v^{\theta},
(\tau  + p) v^{\theta} \right),
\end{eqnarray}
\noindent
\begin{eqnarray}
{\bf S}({\bf w}) = (S_1,S_2,S_3,S_4,S_5),
\end{eqnarray}
with
\begin{eqnarray}
S_1 & = & -{\cal A} D v^r - {\cal B} D v^{\theta},
\\
S_2 & = & -{\cal A} (S_r v^r + p) - {\cal B} S_r v^{\theta}
 + \alpha {\cal C},
\\
S_3 & = & -{\cal A} S_{\theta} v^r - {\cal B} (S_{\theta} v^{\theta} +p)
 + \alpha {\cal D},
\\
S_4 & = & -{\cal A} S_{\phi} v^r - {\cal B} S_{\phi} v^{\theta}
 + \alpha {\cal E},
\\
S_5 & = & -{\cal A} (\tau +p) v^r - {\cal B} (\tau +p) v^{\theta}
 + \alpha {\cal F},
\end{eqnarray}
\noindent
with the definitions
\begin{eqnarray}
{\cal A} &=& \alpha \left( {{r}\over{\varrho^2}} + {{1}\over{2}}
{{\Sigma_{,r}\Delta - \Sigma\Delta_{,r}}\over{\Delta\Sigma}} \right),
\\
{\cal B} &=& \alpha \left( \cot\theta - \frac{a^2}{\varrho^2}\sin\theta\cos\theta
-\frac{a^2\Delta}{2\Sigma}\sin2\theta \right),
\\
{\cal C} &=& T^{rr}g_{rr,r} + T^{r\theta}g_{rr,\theta} 
\nonumber \\
&-& g_{rr} ( T^{tt}\Gamma^{r}_{tt} + T^{rr}\Gamma^{r}_{rr} 
+ T^{\theta\theta}\Gamma^{r}_{\theta\theta} + T^{\phi\phi}\Gamma^{r}_{\phi\phi}
\nonumber \\
&+& 2T^{t\phi}\Gamma^{r}_{t\phi} + 2T^{r\theta}\Gamma^{r}_{r\theta} ),
\\
{\cal D} &=& T^{r\theta}g_{\theta\theta,r} + T^{\theta\theta}g_{\theta\theta,\theta} 
\nonumber \\
&-& g_{\theta\theta} ( T^{tt}\Gamma^{\theta}_{tt} + T^{rr}\Gamma^{\theta}_{rr} 
+ T^{\theta\theta}\Gamma^{\theta}_{\theta\theta} + T^{\phi\phi}\Gamma^{\theta}_{\phi\phi}
\nonumber \\
&+& 2T^{t\phi}\Gamma^{\theta}_{t\phi} + 2T^{r\theta}\Gamma^{\theta}_{r\theta} ),
\\
{\cal E} &=& T^{tr}g_{t\phi,r} + T^{t\theta}g_{t\phi,\theta} +
T^{r\phi}g_{\phi\phi,r} + T^{\theta\phi}g_{\phi\phi,\theta}
\nonumber \\
&-& 2g_{t\phi} ( T^{tr}\Gamma^{t}_{tr} + T^{t\theta}\Gamma^{t}_{t\theta} +
T^{r\phi}\Gamma^{t}_{r\phi} + T^{\theta\phi}\Gamma^{t}_{\theta\phi} )
\nonumber \\
&-& 2g_{\phi\phi} ( T^{tr}\Gamma^{\phi}_{tr} + T^{t\theta}\Gamma^{\phi}_{t\theta} +
T^{r\phi}\Gamma^{\phi}_{r\phi} + T^{\theta\phi}\Gamma^{\phi}_{\theta\phi} ),
\\
{\cal F} &=& T^{tr}\alpha_{,r} + T^{t\theta}\alpha_{,\theta} 
\nonumber \\
&-& 2\alpha (T^{tr}\Gamma^{t}_{tr} + T^{t\theta}\Gamma^{t}_{t\theta} +
T^{r\phi}\Gamma^{t}_{r\phi} + T^{\theta\phi}\Gamma^{t}_{\theta\phi} ).
\end{eqnarray}
\noindent
The `,' in the above expressions denotes partial differentiation and 
the $\Gamma^{\delta}_{\mu\nu}$ stand for the Christoffel symbols (only the
non-vanishing ones are displayed) which are obtained from the metric
according to the usual definition
\begin{eqnarray}
\Gamma^{\alpha}_{\mu\nu} = \frac{1}{2}g^{\alpha\lambda}
\left(\frac{\partial g_{\lambda\nu}}{\partial x^{\mu}} +
\frac{\partial g_{\lambda\mu}}{\partial x^{\nu}} -
\frac{\partial g_{\mu\nu}}{\partial x^{\lambda}} 
\right).
\end{eqnarray}

\section{Numerical method}
\label{numerics}

\subsection{The hydrodynamics code}

The hydrodynamics code used in our computations was originally developed for
studies of relativistic wind accretion on to black holes 
(\cite{font:98a,font:98b,font:98c,font:99}). This code performs the
numerical integration of system (\ref{system}) using a so-called
Godunov-type scheme. Such schemes are specifically designed to solve
nonlinear hyperbolic systems of conservation laws (see, e.g. Toro 1999
for definitions). In a Godunov-type method the knowledge of the
characteristic structure of the equations is essential to design a
solution procedure based upon either exact or approximate Riemann
solvers.  These solvers compute, at every cell-interface of the numerical
grid, the solution of local Riemann problems (i.e., the simplest initial
value problems with discontinuous initial data). Therefore, they
automatically guarantee the proper capturing of all discontinuities
which may arise naturally in the solution space of a nonlinear hyperbolic
system.

The time update of system (\ref{system}) from $t^n$ to $t^{n+1}$ is
performed according to the following conservative algorithm:
\begin{eqnarray}
    {\bf U}_{i,j}^{n+1} = {\bf U}_{i,j}^{n}
    & - & \frac{\Delta t}{\Delta r}
    (\widehat{{\bf F}}^r_{i+1/2,j}-\widehat{{\bf F}}^r_{i-1/2,j})
\nonumber \\
& - & \frac{\Delta t}{\Delta \theta}
    (\widehat{{\bf F}}^{\theta}_{i,j+1/2}-\widehat{{\bf F}}^{\theta}_{i,j-1/2})
\nonumber \\
& + & \Delta t \,\, {\bf S}_{i,j} \, .
\end{eqnarray}
\noindent
Index $n$ represents the time level and the time (space) discretization
interval is indicated by $\Delta t$ ($\Delta r, \Delta\theta$). The
numerical fluxes in the above equation, $\widehat{{\bf F}}^{r}$,
$\widehat{{\bf F}}^{\theta}$ are computed by means of the HLLE Riemann
solver \citep{harten:83,einfeldt:88}. These fluxes are
obtained independently for each direction and the time update of the
state-vector ${\bf U}$ is done simultaneously using a method of lines
in combination with a second-order (in time) conservative Runge-Kutta
scheme. Moreover, in order to set up a family of local Riemann problems
at every cell-interface we use a piecewise linear reconstruction 
procedure \citep{vanleer:79} which provides second-order accuracy in space.

\subsection{Grid and boundary conditions}

We use a computational grid of $300\times 100$ zones in the radial and
angular direction, respectively. The grid is logarithmically spaced in
the radial direction. The innermost radius is located at
$r_\mathrm{min}=2.1$. The location of the maximum radius $r_\mathrm{max}$
depends on the particular model under study. For the stationary models
presented in Section \ref{tests}, we have $r_\mathrm{max}=35.$
Correspondingly, for the simulations of the runaway instability, the
radial grid extends to a sufficiently large distance in order to ensure
that the whole disc is included within the computational domain. The
particular values of $r_\mathrm{max}$ are displayed in
Table \ref{tab:ModelParameters} below. The typical width of the
innermost cell, where we have the highest resolution, is
$\Delta r \simeq 1.9\times 10^{-2}$.

In the angular direction we use a finer grid within the torus and a much
coarser grid outside. The angular zones are distributed according to
the following law:
\begin{eqnarray}
\frac{\theta_{j}}{\pi}\  & = & \left\lbrace\begin{array}{ll}
1.2 \frac{j}{M} & 1 \le j \le \frac{M}{8}\ ,\\
\frac{1}{3}\left(0.1+2.8\frac{j-1}{M}\right) & \frac{M}{8}+1\le j \le \frac{7M}{8}+1\ ,\\
-0.2+1.2\frac{j-1}{M} & \frac{7M}{8}+1\le j \le M+1\ .\\
\end{array}\right.
\end{eqnarray}

Although the flows we are simulating have equatorial plane symmetry
we extend the computational domain in the angular direction from 0 to
$\pi$. This allows us to measure the ability of the code in keeping a
symmetric evolution. 

For the second-order numerical scheme we use we need to impose boundary 
conditions in two additional zones at each end of the domain. The boundary 
conditions are applied to $\rho, v_r, v_{\theta}$ and $v_{\phi}$ and they 
are as follows: at the inner boundary $r_{\mathrm{min}}$ all velocities are 
linearly extrapolated to the boundary zones from the innermost zones in the 
physical grid. The density is however assumed to have zero gradient across 
the inner boundary. The rest of thermodynamical quantities are computed using
the polytropic EoS. At the outer radial boundary $r_{\mathrm{max}}$ all
variables keep the constant initial values given by the choice of the
particular disc solution. Reflection boundary conditions are used at both
poles ($\theta=0$ and $\pi$), i.e., all variables are symmetric, except for
$v_\theta$ which changes sign. 

\subsection{Additional aspects}

Since we are considering adiabatic evolutions, we only solve
for the first four equations of system (\ref{system}). The internal
energy (proportional to the rest-mass density) is obtained algebraically
using a polytropic EoS, $p=\kappa\rho^{\gamma}$, i.e., 
$\varepsilon=\frac{\kappa}{\gamma-1}\rho^{\gamma-1}$. After the time update
of the conserved quantities, the primitive variables are recomputed. As the 
relation between the two sets of variables is not in closed algebraic 
form, the primitive variables are computed using the following procedure: 
The evolved quantities $D$ and $S_{i}$ being known, we eliminate $\rho$ and
$h$ from the definition of $S_{i}$ given by Eq.~(\ref{eq:DefLabFrame}) to
express the norm $S^{2}=\left(\rho h\right)^{2}\Gamma^{4}v^{2}$ of $S_{i}$
as a function of the Lorentz factor $\Gamma$ only:
\begin{equation}
S^{2}(\Gamma) = D^{2}
\left(1+\frac{\gamma}{\gamma-1}\kappa\left(\frac{D}{\Gamma}\right)^{\gamma-1}\right)^{2}
\left(\Gamma^{2}-1\right)\ .
\end{equation}
We solve this equation by an iterative Newton-Raphson algorithm. Once the
Lorentz factor $\Gamma$ is found, the other primitive variables are easily
derived using the relations $\rho=D/\Gamma$,
$h=1+\frac{\gamma}{\gamma-1}\kappa\rho^{\gamma-1}$ and
$v_{i}=S_{i}/\left(\rho h \Gamma^{2}\right)$.

Finally, it is worth pointing out that in order to evolve the ``vacuum"
zones which lie outside the disc using a hydrodynamics code, we adopt
the following simple yet effective procedure. Before constructing the
initial torus we build up a background spherical accretion solution of
a sufficiently low density so that its presence does not affect the
dynamics of the disc. This stationary solution is given by the relativistic
extension of the spherical Bondi accretion solution derived by \citet{michel:72}.
This solution depends on the location of the critical point
$r_\mathrm{c}$ and of the density at this point $\rho_\mathrm{c}$,
together with the adiabatic exponent and polytropic constant of the EoS,
which we chose the same as inside the torus. In our approach we
chose the values of $r_{c}$ and $\rho_{c}$ (which is computed from
$r_\mathrm{c}$ with the condition that the flow is regular at the
critical point) in order to impose that the maximum density in the
background spherical solution is $5.0\times 10^{-6}$ times the maximum
density at the center of the disc. By doing this we have checked that
the rest-mass present in our background solution is always negligible
compared to the mass of the disc and that the associated mass flux corresponding 
to this spherical accretion is also negligible compared to the mass flux 
from the disc. We note that in the outermost part the values of the background 
density can be as low as $10^{-8}$ times the maximum density at the center 
of the disc.

\section{Simulations of stationary models}
\label{tests}

As mentioned in the previous section our hydrodynamics code has been used 
previously in a number of relativistic wind accretion simulations. However, 
in order to test the ability of the code when dealing with accretion discs 
we have first considered time-dependent simulations of stationary models. 
The aim of these simulations has been to find out whether the code is 
capable of keeping those models in equilibrium during a sufficiently long 
period of time (much larger than the rotation period of the disc). In order
to do so we have considered the same stationary models that
\cite{igumenshchev:97} analyzed, in the limit of no black hole rotation.
These four models are characterized by $l=3.9136$, a value in between the
marginally stable and marginally bound orbits, and an increasingly large
value of the energy gap at the cusp, $\Delta W_\mathrm{in} = 0.02$, $0.04$,
$0.08$ and $0.16$. Similarly, the polytropic EoS has been chosen with an
adiabatic index $\gamma=4/3$ and a polytropic constant $\kappa =
1.5\times 10^{20}\ \mathrm{cgs}$. Furthermore, the mass of the black hole 
is kept constant throughout these test evolutions.

\begin{figure}
\begin{center}
\begin{tabular}{c}
\psfig{file=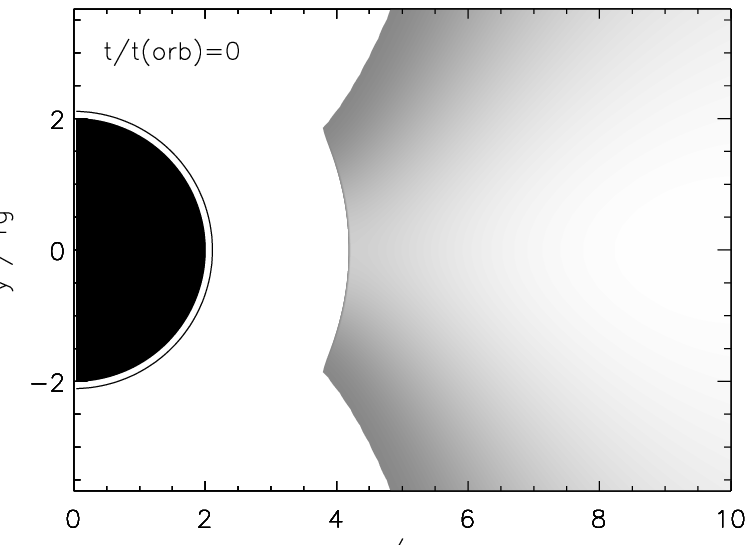,width=8cm}\\
\vspace*{1ex}
\\
\psfig{file=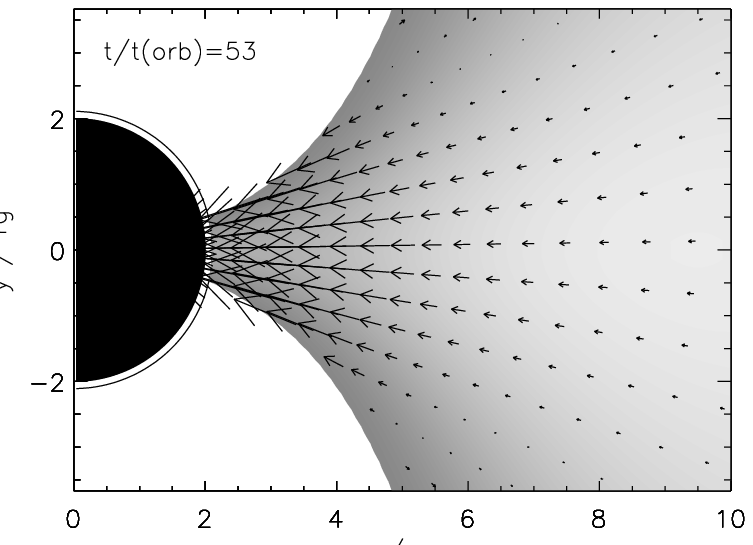,width=8cm}\\
\end{tabular}
\end{center}
\caption{Morphology of the inner part of the disc with parameters $l=3.9136$ 
and $\Delta W_\mathrm{in}=0.16$. \textit{Top:} initial state; \textit{Bottom:} 
final state at $t \simeq 53 t_\mathrm{orb}$. The arrows are proportional to the 
components $\left( S_{r} / \sqrt{g_{rr}}, S_{\theta}/\sqrt{g_{\theta\theta}}\right)$ 
of the momentum and are plotted only in the region where $\rho \ge 0.05\ \rho^{0}_\mathrm{max}$. 
The black hole is represented by the black circle and the exterior circle around it 
marks the location of the inner boundary of the grid.}
\label{discmorphology}
\end{figure}

\begin{figure}
\centerline{\psfig{file=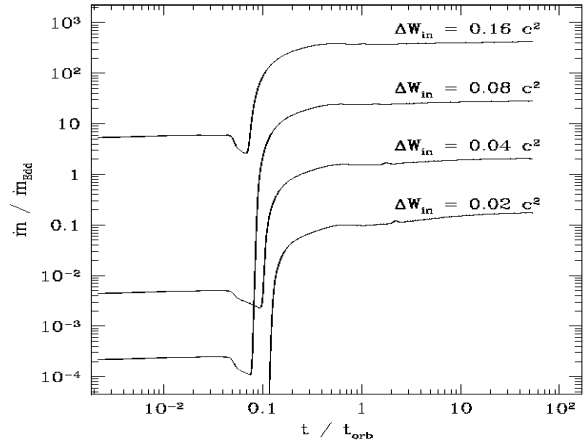,width=8.0cm,height=6.0cm}}
\caption{Time evolution of the mass flux for a stationary model with
$l=3.9136$ and $\Delta W=0.02$, $0.04$, $0.08$ and $0.16$. The time is
given in units of the orbital period at the center (see text), which
here is $t_\mathrm{orb}\simeq 187\ r_\mathrm{g}/c$. The mass flux is
normalized with the Eddington limit computed with $M_\mathrm{BH}=1\
\mathrm{M}_\odot$. For a different choice of polytropic constant
$\kappa$ and black hole mass $M_\mathrm{BH}$, the mass flux scales
as $\kappa^{-3} M_\mathrm{BH}$ for $\gamma=4/3$. The initial value
of the mass flux is fixed by the spherical mass accretion rate
associated with the Bondi flow imposed in the low density regions
outside the torus.  Notice that after a transition phase lasting for 
$\sim 0.1\ t_\mathrm{orb}$ the mass accretion rate rapidly tends,
asymptotically, to the stationary values.}
\label{mdotvstime}
\end{figure}

As a representative example Figure \ref{discmorphology} shows the
morphology of the model with $\Delta W_\mathrm{in} = 0.16$. The cusp
is located at $r_\mathrm{cusp} = 4.197$ and the center at
$r_\mathrm{center}=9.598$. The grid extends to $r_\mathrm{max}=35$.
The top panel shows a gray-scale plot of the logarithm of the density for
the initial model at $t=0$, together with the corresponding velocity
field. The bottom panel shows the same morphology at the final time
$t=10^4$. This corresponds to about 53 dynamical timescales,
choosing as a dynamical timescale the orbital period
$t_\mathrm{orb}=2\pi/\Omega$ at $r=r_\mathrm{center}$, which
for this model is $t_\mathrm{orb}=184.9$. The code was stopped after
roughly $2\times 10^5$ iterations with no signs of numerical instabilities
present. One can clearly see in Figure \ref{discmorphology} that an accretion 
flow from the disc to the black hole appears in the inner region of the grid. 
This flow, however, becomes rapidly stationary (see below). In addition, for 
our particular choice of model parameters, the corresponding mass flux is 
very low. Therefore, except in the innermost region where the accretion 
flow develops, the morphology of the disc remains essentially unchanged 
during the whole evolution from $t=0$ to $t=53 t_\mathrm{orb}$.

There are some additional noteworthy issues concerning this figure:
first, it clearly shows the ability of the code to keep the equatorial 
plane symmetry of the torus, even though the angular domain extends from
$0$ to $\pi$. Second, the smoothness of the initial disc distribution in the 
grid is maintained during the whole evolution even close to the black hole
horizon. Finally, contrary to previous work 
\citep{hawley:84b,igumenshchev:97} there are no hints of vortices developing
in the flow. In \citet{hawley:84b} such vortices are associated 
to small poloidal velocities and are not as noticeable as in the results
of \citet{igumenshchev:97}. The latter claim, however, that such
vortex motions are likely to be related to the choice of initial
conditions, developing from initial perturbations close to the cusp 
and propagating outwards undamped.

A more quantitative proof of the ability of the code in keeping the
stationarity of this solution is provided in Figure~\ref{mdotvstime}.
This figure shows the evolution of the mass accretion rate (normalized
to the Eddington value), $\dot{m}/\dot{m}_\mathrm{Edd}$, as a function
of $t / t_\mathrm{orb}$. The mass flux is computed at the innermost
radial point as
\begin{eqnarray}
\dot{m}= 2\pi\int_0^{\pi} \sqrt{-g} D v^r d\theta,
\end{eqnarray}
where the volume element for the Schwarzschild metric is given by
$\sqrt{-g}=r^2\sin\theta$. The Eddington mass flux $\dot{m}_\mathrm{Edd}=
\frac{L_\mathrm{Edd}}{c^{2}}\simeq 1.4\ 10^{17}\ \frac{M_\mathrm{BH}}
{\mathrm{M}_\odot}\ \mathrm{g/s}$ is computed for $M_\mathrm{BH}=1\
\mathrm{M}_\odot$. Rescaling of $\dot{m}$ for different polytropic
constant $\kappa$ and black hole mass $M_\mathrm{BH}$ is given by
\begin{equation}
\frac{\left(\dot{m}/\dot{m}_\mathrm{Edd}\right)_{1}}
{\left(\dot{m}/\dot{m}_\mathrm{Edd}
\right)_{2}} = \left(\frac{\kappa_{1}}{\kappa_{2}}\right)^{-3}
\frac{M_{\mathrm{BH}\ 1}}
{M_{\mathrm{BH}\ 2}}\ ,
\end{equation}
where $\gamma=4/3$ was assumed. After a transient initial phase the
mass accretion rate is seen to rapidly tend, asymptotically, to a constant value. 
The offset observed during the initial phase corresponds to the spherical 
accretion mass flux associated with the particular background solution we 
use outside the torus. 

\begin{figure}
\centerline{\psfig{file=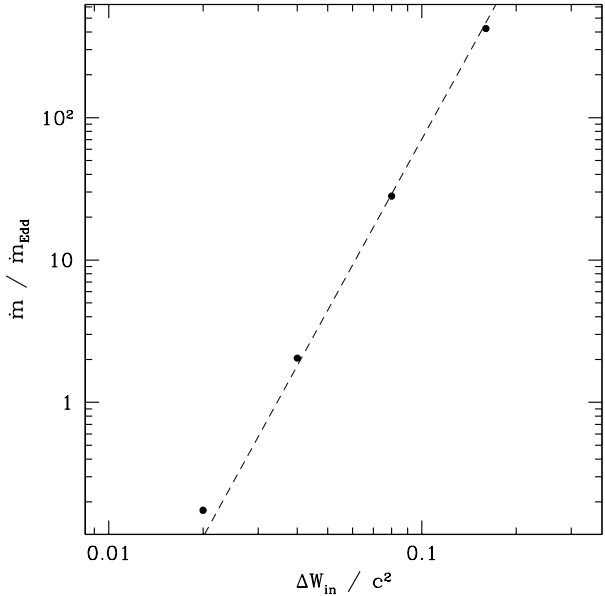,width=8.5cm}}
\caption{Mass flux $\dot{m}$ as a function of the energy gap
$\Delta W_\mathrm{in}$ between the inner edge of the disc and the cusp.
The dots indicate the asymptotic values we get in our simulations
(see Fig.~\ref{mdotvstime}). The dashed line shows the slope which is
expected from the theoretical prediction given by Eq.~(\ref{eq:mdotvsdw}),
which for $\gamma=4/3$ is 4. The plot corresponds to simulations with
$M_\mathrm{BH}=1\ \mathrm{M}_\odot$ and $\kappa = 1.5 \times 10^{20}\
\mathrm{cgs}$.}
\label{massfluxvsdw}
\end{figure}

\begin{figure}
\begin{tabular}{c}
\psfig{file=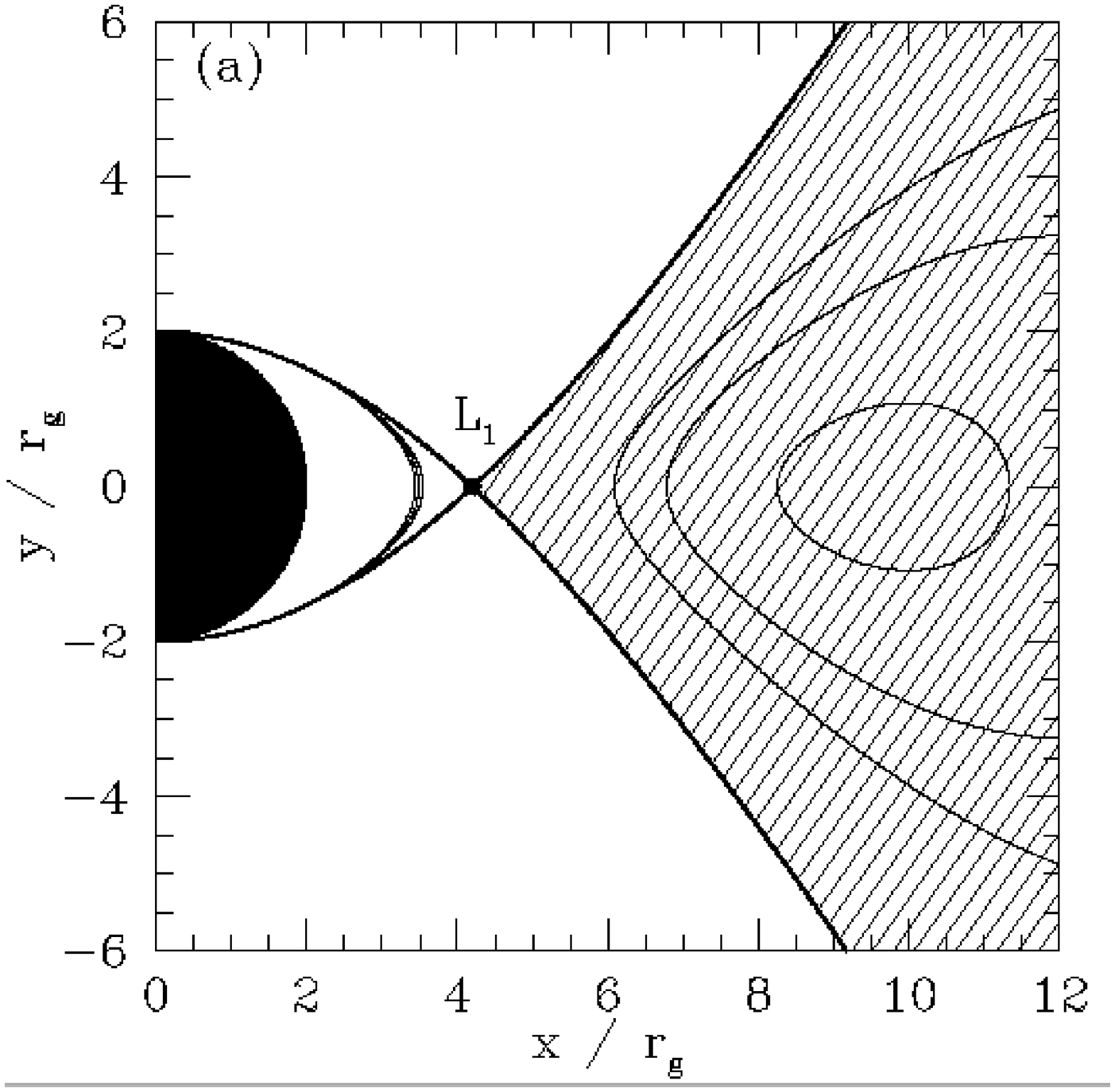,width=6.65cm}\\
\psfig{file=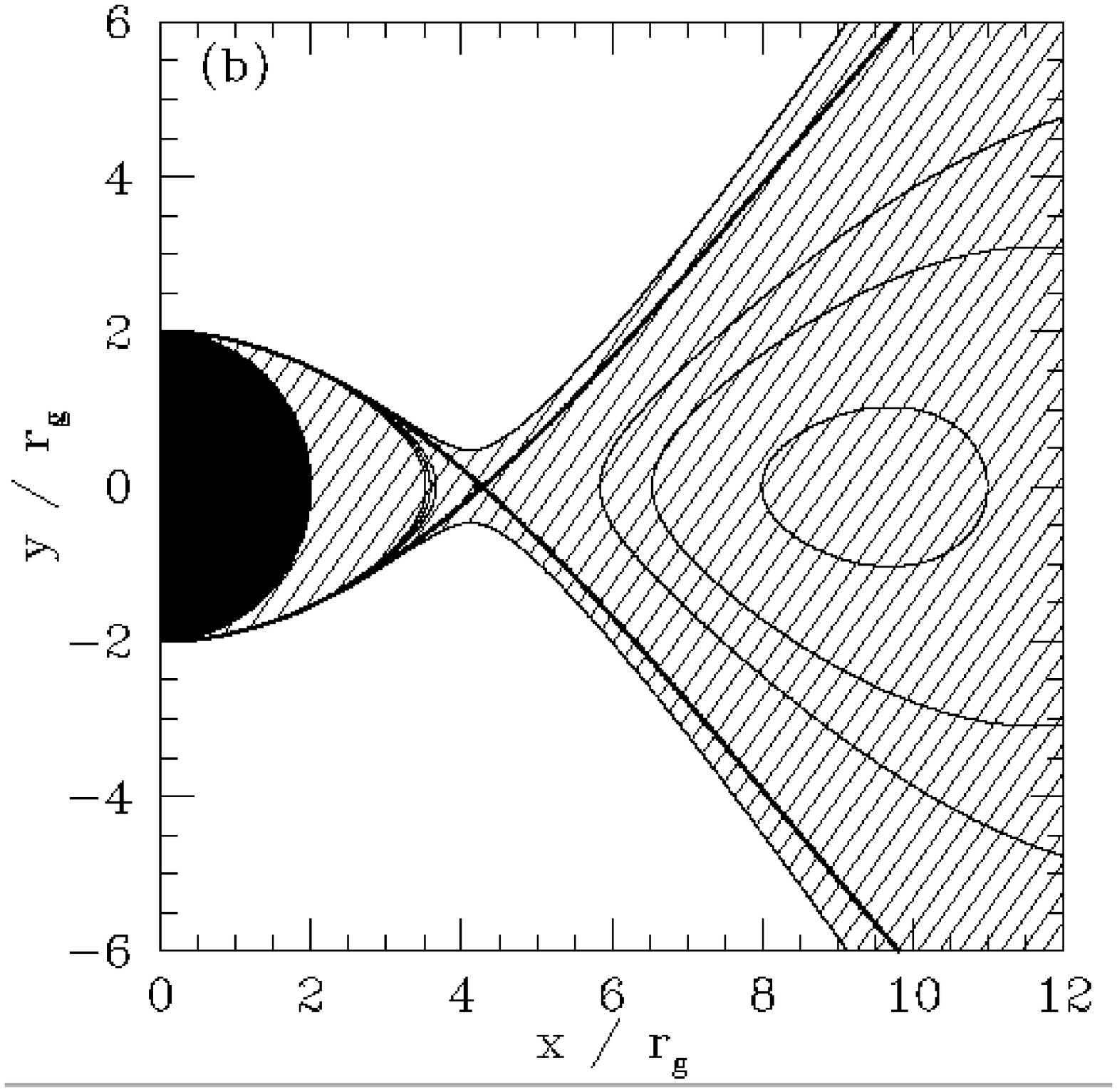,width=6.65cm}\\
\psfig{file=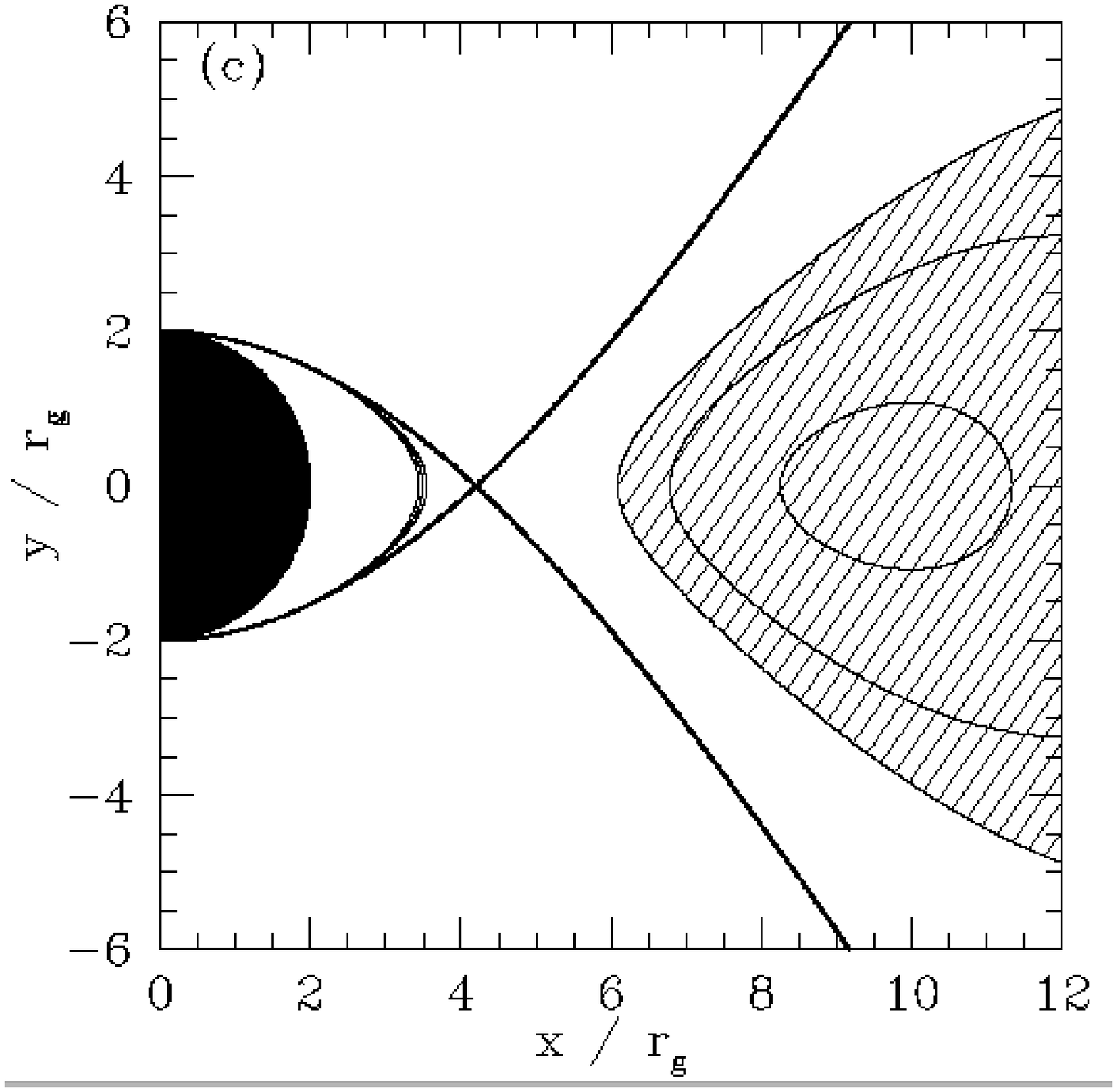,width=6.65cm}\\
\end{tabular}
\caption{\textbf{The physical origin of the runaway instability:} the initial disc 
fills its Roche lobe (panel (a)). Mass transfer from the disc to the black hole 
occurs through the cusp, which is located at the Lagrange point $L_{1}$. The mass 
of the black hole increases and the disc has to find a new equilibrium configuration 
with the new gravitational potential. There exist two possibilities: (i) the disc 
overflows its Roche lobe (panel (b)). This process speeds up the mass transfer 
and the disc becomes \textit{unstable}. (ii) The disc contracts inside its Roche lobe 
(panel (c)), which forces the mass transfer to slow down, resulting on a \textit{stable} disc.}
\label{fig:TheoryRunaway}
\end{figure}

Next, in Figure~\ref{massfluxvsdw} we plot the mass flux as a function of 
the energy gap $\Delta W_\mathrm{in}$ for the four stationary models
we have considered. The values selected for the mass flux in each model
are the asymptotic ones, obtained after the simulations have been evolved
up to a final time $t=10^4$, roughly 54 orbital periods. This plot
allows us to check if the code is able to reproduce the analytic
dependence given by Eq.~(\ref{eq:mdotvsdw}). For a $\gamma=4/3$
polytrope the expected slope is 4 (dashed line). As the figure shows
our results are in good agreement with this analytic prediction as well
as with the numerical results obtained by~\cite{igumenshchev:97} for the
same models.

\section{Simulations of the runaway instability}
\label{simulations}

\subsection{The physical origin of the instability}

\begin{figure}
\centerline{\psfig{file=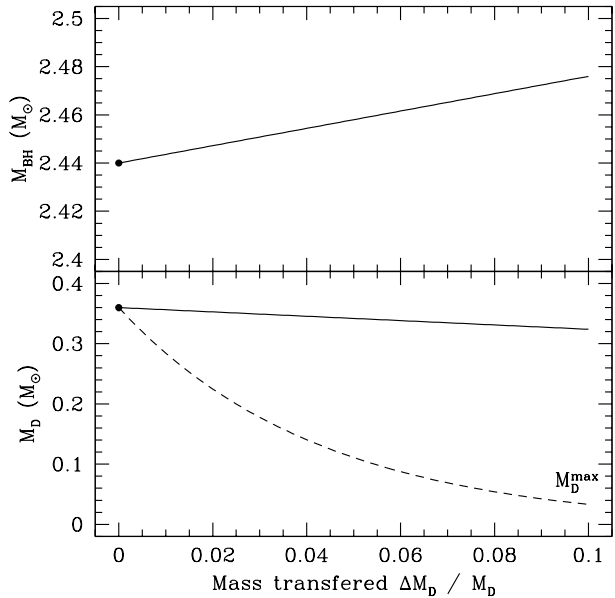,width=8.5cm}}
\caption{\textbf{The physical origin of the runaway instability:} the evolution of 
models with different mass is plotted against the mass transfered from the disc to 
the black hole. The initial disc is filling its Roche lobe (indicated by the big 
dot symbol). Initially $M_{BH}=2.44\ \mathrm{M}_\odot$ and $M_{D}=0.36\ 
\mathrm{M}_\odot$. \textit{Upper panel:} mass $M_\mathrm{BH}$ 
of the black hole; \textit{Lower panel :} mass $M_\mathrm{D}$ of the disc and 
mass $\mathrm{M}_\mathrm{D}^\mathrm{max}$ contained inside the Roche lobe 
(dashed line). As $M_\mathrm{D} \ge M_\mathrm{D}^\mathrm{max}$, the disc 
overflows its Roche lobe which speeds up the mass transfer and leads to the 
runaway instability.}
\label{fig:RunawaySchwarzschild}
\end{figure}

The physical mechanism leading to the runaway instability has been explained by 
\citet{abramowicz:83} and \citet{nishida:96a}. The mass transfer from the disc 
to the black hole starts once the disc fills its Roche lobe 
(see Fig.~\ref{fig:TheoryRunaway}a). From this moment any small perturbation allows 
the gas to flow through the cusp located at the inner edge of the disc. As a result 
the mass of the black hole increases, the equipotential surfaces move, and the radial 
location of the cusp changes. The disc has to find a new equilibrium configuration: 
one possibility is that the disc overflows its Roche lobe as is depicted in 
Fig.~\ref{fig:TheoryRunaway}b. In this case the mass transfer speeds up, which 
leads to the runaway instability. Alternatively, the disc may contract inside 
its Roche lobe (see Fig.~\ref{fig:TheoryRunaway}c). In this case the mass transfer 
slows down. The mass flux is self-regulated by this process and the accretion is stable. 

However, for a Schwarzschild black hole and a constant angular momentum distribution 
in the disc, most discs are unstable \citep{abramowicz:83} (see also Table~\ref{tab:studies}). 
This is illustrated in Fig.~\ref{fig:RunawaySchwarzschild} for the particular case 
where the black hole mass is $M_\mathrm{BH}=2.44\ \mathrm{M}_\odot$, the disc mass 
is $M_\mathrm{D}=0.36\ \mathrm{M}_\odot$, the adiabatic index is $\gamma=4/3$ and 
the polytropic constant is $\kappa=4.76 \times 10^{14}\ \mathrm{cgs}$ (which corresponds 
to degenerate relativistic electrons with $Y_\mathrm{e}=0.5$ electrons per nucleon). 
In this figure the evolution of models with different mass is plotted against the 
mass transfered from the disc to the black hole. The initial disc is filling its 
Roche lobe. The location of both the initial disc mass and black hole mass is 
indicated by a big dot symbol in each panel. The upper panel shows the evolution 
of the mass of the black hole whereas the lower one shows the corresponding evolution 
of the disc and of $\mathrm{M}_\mathrm{D}^\mathrm{max}$, which is the maximum disc 
mass contained inside the Roche lobe (dashed line). As $M_\mathrm{D} \ge 
M_\mathrm{D}^\mathrm{max}$, the disc overflows its Roche lobe which speeds up the mass 
transfer and leads to the runaway instability.

\subsection{The sequence of stationary metrics approximation}
\label{sec:MethodBH}

The increase of the mass of the black hole is the fundamental process triggering 
the runaway instability. In order to properly take into account the dynamical 
evolution of the underlying gravitational field, one should solve the coupled 
system of Einstein and hydrodynamic equations. However, such a task has not yet 
been accomplished in the context of the runaway instability
and, to some extent, it may be still far from the capabilities 
of current codes in numerical relativity. Numerical stability considerations, 
coming from both the coordinate singularity existing at the rotation axis 
($\theta=0, \pi$), which spoils the long-term integration of the Einstein equations 
even in vacuum~\citep{brandt:95}, and from the mathematical formulation of the 
field equations themselves, make such a task very challenging. Numerical 
relativity codes evolving black hole spacetimes with perfect fluid matter are 
only becoming available very recently, both in axisymmetry~\citep{brandt:00} 
as in full 3D~\citep{shibata:00,shibata2:00,alcubierre:00,font:01}. 
Nevertheless, the resolution and the 
integration times required to study the time evolution of the runaway instability 
may be still too demanding for such relativistic codes which incorporate self-gravity.

For all these reasons and for the complete lack of time-dependent simulations of 
the runaway instability in relativity, we have adopted a simplified and pragmatic 
approach to the problem. In our procedure the spacetime metric is approximated at 
each time step by a stationary exact black hole metric of varying mass (and angular 
momentum in the case of a rotating black hole). The mass $M$ of the black hole 
necessary to compute the metric coefficients is increased at each time step $\Delta t$ 
according to:
\begin{equation}
M^{n+1} =  M^{n}+\Delta t\ \dot{m}^{n}\ ,
\label{massincrease1}
\end{equation}
where the mass flux at the inner radius of the grid is evaluated by the equation
\begin{equation}
\dot{m}^{n} = 2\pi r_{i_\mathrm{min}}^{2} \Sigma_{j=1}^{j_{\mathrm{max}}} 
\left(\theta_{j+1}-\theta_{j}\right) \sin\theta_{j} D_{i_\mathrm{min} j} 
v^{r}_{i_\mathrm{min} j}.
\label{massincrease2}
\end{equation}

As the mass of the black hole increases during the simulations, the horizon moves 
outwards. To avoid the inner radius of the grid to become smaller than the radius 
of the growing horizon, we increase the index $i_\mathrm{min}$ of the first radial 
zone when necessary, so that the condition $r_{i_\mathrm{min}} > 2M$ is always 
respected. We notice that the black hole mass increases very slowly during the
evolution which implies that the metric coefficients at any time step differ very 
little from the final values which would correspond to an exact
Schwarzschild black hole of bigger mass but with no matter around.

\subsection{Initial state}

A given initial state of the black hole plus disc system is determined by five 
parameters: the mass of the black hole $M_\mathrm{BH}$, the specific angular 
momentum in the disc $l$, the potential at the inner edge of the disc 
$\Delta W_\mathrm{in}$, the polytropic constant $\kappa$ and the adiabatic index $\gamma$. 
The computing time needed for one hydrodynamical simulation is too large to allow 
for a complete exploration of this parameter space. For this reason we focus on 
those models which are expected to be found in the central engine of GRBs. These 
systems are formed either after the coalescence of two compact objects or after 
the gravitational collapse of a massive star.

\begin{table*}
\begin{center}
\caption{Initial models. The following parameters are listed: mass of the black 
hole $M_\mathrm{BH}$, disc-to-hole mass ratio $M_\mathrm{D}/M_\mathrm{BH}$, 
specific angular momentum in the disc $l$, potential barrier at the inner 
edge $\Delta W_\mathrm{in}$, mass flux in the stationary regime $\dot{m}_\mathrm{stat}$, 
minimum and maximum radii of the grid, 
$r_\mathrm{min}$ and $r_\mathrm{max}$, radius of the cusp $r_\mathrm{cusp}$, 
radius of the center $r_\mathrm{center}$ (all radii are in units of the gravitational 
radius $r_\mathrm{g}$), and orbital period at the center of the disc $t_\mathrm{orb}$. 
The last column lists the timescale associated with the runaway instability 
as defined in Section \ref{sec:ResultsUnstable}. In all cases, the EoS parameters
are $\kappa = 4.76\times 10^{14}\ \mathrm{cgs}$ and $\gamma=4/3$.}
\label{tab:ModelParameters}
\begin{tabular}{l|cc|ccc|cc|ccc|c}
\hline
Model & $M_\mathrm{BH}$ & $M_\mathrm{D}/M_\mathrm{BH}$ & $l$ & 
$\Delta W_\mathrm{in}$ & $\dot{m}_\mathrm{stat}$ 
& $r_\mathrm{min}$ & $r_\mathrm{max}$ & $r_\mathrm{cusp}$ & $r_\mathrm{center}$ & 
$t_\mathrm{orb}$  & $t_\mathrm{run}/t_\mathrm{orb}$\\
&  $(\mathrm{M}_\odot)$ & & & & $(\mathrm{M_{\odot}/s})$ & & & & (geo/ms) & \\
\hline
1a & 2.5 & 1.   & 3.9325 & 0.005 & 0.090 & 2.1 & 85.$^{*}$ & 4.1492 & 9.7930 & 193. / 2.37 & 97. / $95.^{\dagger}$\\ 
2a & 2.5 & 1.   & 3.9085 & 0.01  & 0.28  & 2.1 & 85.$^{*}$ & 4.2105 & 9.5455 & 185. / 2.28 & 38. / 36.$^{\dagger}$\\ 
3a & 2.5 & 1.   & 3.8564 & 0.02  & 2.1   & 2.1 & 85.$^{*}$ & 4.3639 & 8.9919 & 169. / 2.09 & 10. / 9.2$^{\dagger}$\\ 
4a & 2.5 & 1.   & 3.7255 & 0.04  & 34.   & 2.1 & 85.$^{*}$ & 5.0104 & 7.3644 & 126. / 1.55 & 3.8 / 3.3$^{\dagger}$\\
\hline
1b & 2.5 & 0.1  & 3.8749 & 0.005 & 0.032 & 2.1 & 37.       & 4.3058 & 9.1915 & 175. / 2.16 & 140.\\
2b & 2.5 & 0.1  & 3.8459 & 0.01  & 0.17  & 2.1 & 37.       & 4.3990 & 8.8768 & 166. / 2.05 & 64. \\
3b & 2.5 & 0.1  & 3.7798 & 0.02  & 1.9   & 2.1 & 37.       & 4.6698 & 8.1077 & 145. / 1.79 & 12. \\ 
\hline
1c & 2.5 & 0.05 & 3.8798 & 0.001 & 0.21  & 2.1 & 32.       & 4.2911 & 9.2438 & 177. / 2.17 & 110. \\
\hline
\end{tabular}
\end{center}
\begin{tabular}{l}
$^{*}$ In models 1a to 4a the grid consists of a first grid from $r=2.1$ to 
$r=28$ and a second grid from $r=28$ to $r=85$ to 
avoid $\Delta r$ at\\ the inner radius becoming too large.\\
$^{\dagger}$ The second estimate of $t_\mathrm{run}$ given for models 1a to 4a
corresponds to simulations where the mass of the black hole starts to increase\\ only once
the stationary regime has been reached (see Section~\ref{sec:ResultsUnstable}).
\end{tabular}
\end{table*}

\subsubsection{Black hole mass and disc-to-hole mass ratio} 

As it has been shown by numerical simulations using Newtonian and post-Newtonian 
gravity, both the coalescence of two neutron stars \citep{ruffert:99} and the merger 
of a black hole and a neutron star \citep{kluzniak:98} lead to the formation of 
comparable systems, where the mass of the central black hole and the disc-to-hole 
mass ratio are respectively $M_\mathrm{BH} \sim 2.5\ \mathrm{M_{\odot}}$ and 
$M_\mathrm{D}/M_\mathrm{BH}\sim 0.04$--$0.08$ for the first case, and 
$M_\mathrm{BH}\sim 3\ \mathrm{M_{\odot}}$ and $M_\mathrm{D}/M_\mathrm{BH}\sim 0.2$ 
in the second case. More recently, the fully relativistic 
simulations of binary neutron star coalescence performed by \citet{shibata:00} yield 
disc masses of $\sim 0.05-0.1 M_{*}$ for corotational binaries and $<0.01 M_{*}$ for 
irrotational binaries, where $M_{*}$ is the total rest-mass of the system (typically 
$\sim 2M_{\odot}$).

The case of collapsars is more complex since, in particular, the mass of the disc 
goes on increasing after the formation of the black hole, as matter from the outer 
parts of the star is still infalling. The simulations performed by \citet{macfadyen:99} 
and \citet{aloy:00} start from the $14\ \mathrm{M_{\odot}}$ helium core of a rotating 
$35\ \mathrm{M_{\odot}}$ main sequence star which collapses to produce a central black 
hole surrounded by a disc. When this systems enters a quasi-steady state, the mass of 
the black hole is $2$--$3\ \mathrm{M_{\odot}}$ and the disc-to-hole mass ratio is 
typically about $0.001$--$0.01$.

Taking into account these various results, we have decided to fix the mass of the black 
hole to $M_\mathrm{BH}=2.5\ \mathrm{M_{\odot}}$ and to adjust the angular momentum 
$l$ to get realistic disc-to-hole mass ratios. We have considered three possible 
ratios: $0.05$, $0.1$ and $1$. Such values are very close to what is obtained in 
the simulations of binary coalescence and not too far from the results of the simulations 
of collapsars.

\subsubsection{Equation of state} 

We fix the adiabatic index to $\gamma=4/3$ and the polytropic constant to 
$\kappa=1.2\times 10^{15}\ Y_\mathrm{e}^{4/3}$ with $Y_\mathrm{e}=0.5$. This 
corresponds to an EoS dominated by the contribution of relativistic degenerate electrons 
(the typical density in the disc is $\sim 10^{11}$-$10^{12}\ \mathrm{g/cm^{3}}$). 
We note that such simplified EoS is nevertheless adequate to our purpose since
the work of \cite{nishida:96b} showed that, for stationary models, the effects of 
realistic EoS on the stability of constant angular momentum discs is negligible,
the discs being unstable in all cases.

\subsubsection{Mass flux} 

Once $M_\mathrm{BH}$, $M_\mathrm{D}/M_\mathrm{BH}$, $\kappa$ and $\gamma$ have 
been specified, the only remaining parameter is the potential at the inner edge 
of the disc, $\Delta W_\mathrm{in}$. We choose its value so that the corresponding 
mass flux in the stationary regime (as described in Section \ref{tests}) explores 
a realistic range. In the simulations of binary neutron star coalescence 
carried out by \citet{ruffert:99}, the mass accretion rate of the black hole varies 
between $1$ and $10^{4}\ \mathrm{M_{\odot}/s}$. On the other hand, \citet{kluzniak:98} 
find comparable values in the case of a neutron star - black hole merger. In their 
simulations of collapsars, \citet{macfadyen:99} find a typical mass flux of 
$0.6$--$0.8\ \mathrm{M_\mathrm{\odot}/s}$.

Such mass fluxes are many orders of magnitude larger than the Eddington limit, 
which is $1.2\times 10^{-16}\ \mathrm{M_{\odot}/s}$ for a $M_\mathrm{BH} = 2.5\ 
\mathrm{M}_{\odot}$. However such very high mass fluxes are precisely what is 
required to explain the observed luminosity of GRBs, which is typically 
$L_\mathrm{\gamma} = 10^{51}\ L_{51}\ \mathrm{erg/s}$ in gamma-rays. If this 
radiation is due to internal shocks propagating within an ultra-relativistic wind, 
the kinetic energy flux of this wind is $L_\mathrm{kin} = L_\mathrm{\gamma}/f_\mathrm{\gamma}$, 
where $f_\mathrm{\gamma}$ is the efficiency of the kinetic energy to radiation conversion. 
If one assumes that the production of the relativistic wind is accretion-powered with 
an efficiency $f_\mathrm{acc}$, then the mass flux is given by
\begin{equation}
\dot{m} = \frac{1}{f_\mathrm{acc}}\frac{L_\mathrm{kin}}{c^{2}} = 
0.2\ \left(\frac{f_\mathrm{acc}}{0.05}\right)^{-1} 
\left(\frac{f_\mathrm{\gamma}}{0.05}\right)^{-1} L_{51}\ \mathrm{M_{\odot}/s}\ .
\end{equation}
This estimate is of course no longer relevant if the main energy reservoir powering 
the burst is the rotational energy of the black hole, which can be extracted by the 
Blandford-Znajek effect \citep{blandford:77}.

Taking into such high mass fluxes we have considered the following values of 
$\Delta W_\mathrm{in}$: 0.001, 0.005, 0.01, 0.02 and 0.04 so that the mass flux 
of our initial models in the stationary regime spans $\sim 0.03$--$30\ 
\mathrm{M_{\odot}}/s$. From all the above considerations we have prepared eight 
initial models which are summarized in Table~\ref{tab:ModelParameters}.

\subsection{Results}
\label{sec:ResultsUnstable}

Each of our eight initial models is evolved twice: 
in the first series of runs we keep constant the mass of the black hole while 
in the second set of evolutions such mass is allowed to increase according to 
the law specified in the preceding section, Eqs.~(\ref{massincrease1}) and 
(\ref{massincrease2}). For the first type of runs the mass flux rapidly reaches 
a stationary regime (like in the models discussed in Section \ref{tests}) which 
can be maintained for as long as desired. In practice the final time corresponds 
to many orbital periods of the disc. The mass flux at this stage has the value 
reported as $\dot{m}_\mathrm{stat}$ in Table \ref{tab:ModelParameters}.
On the other hand, when the mass of the black hole increases (second series of runs), 
the time evolution of the system changes dramatically and the runaway instability appears.

\begin{figure}
\centerline{\psfig{file=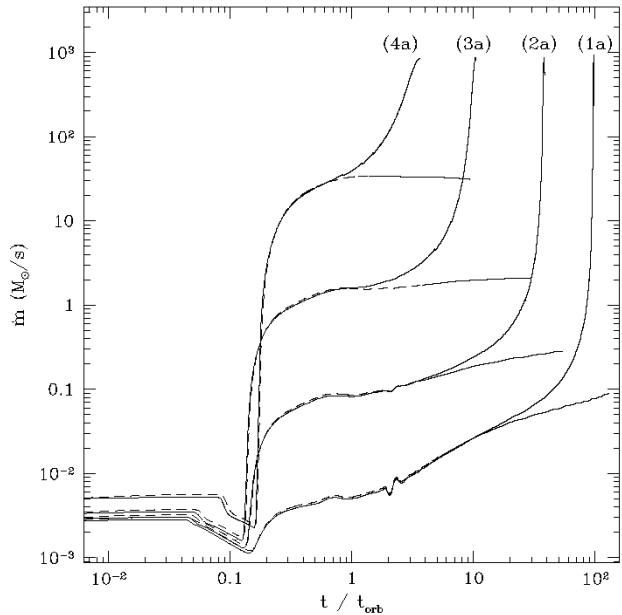,width=\hsize}}
\caption{
Time-evolution of the mass flux for models
(1a) to (4a) of Table \ref{tab:ModelParameters}. The solid lines correspond to
evolutions in which the black hole mass increases with time
according to the procedure explained in Section \ref{sec:MethodBH}.
For comparison, the mass flux in the stationary regime (when the mass of 
the black hole is kept constant) is also plotted using dashed lines. As expected, 
for a black hole of growing mass the accretion process becomes rapidly unstable. 
Notice how the mass flux diverges.}
\label{fig:Instability1}
\end{figure}

\begin{figure}
\centerline{\psfig{file=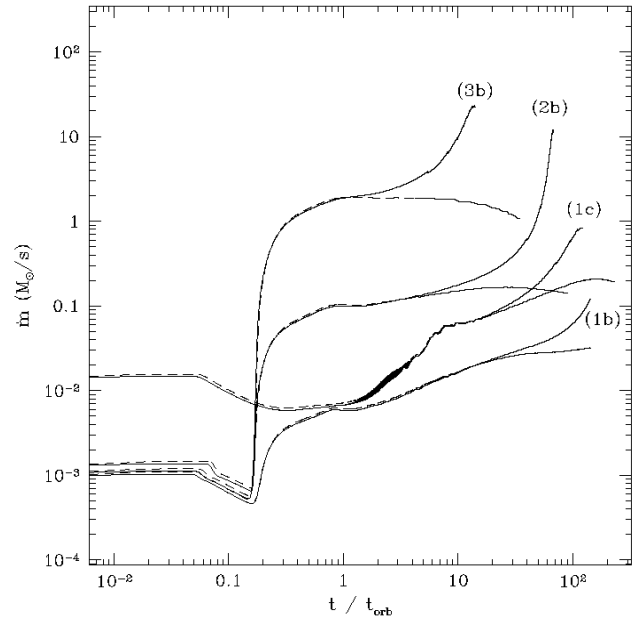,width=\hsize}}
\caption{
Time-evolution of the mass flux for models
(1b) to (3b) and model (1c) of Table \ref{tab:ModelParameters}. As in 
Fig.~\ref{fig:Instability1}, the solid lines correspond to
evolutions in which the black hole mass increases with time
according to the procedure explained in Section \ref{sec:MethodBH}
and the dashed lines correspond to evolutions in which the mass
of the black hole is kept constant. As expected, for a black hole 
of growing mass the accretion process becomes rapidly unstable. 
Notice how the mass flux diverges.}
\label{fig:Instability2}
\end{figure}

\begin{figure}
\centerline{\psfig{file=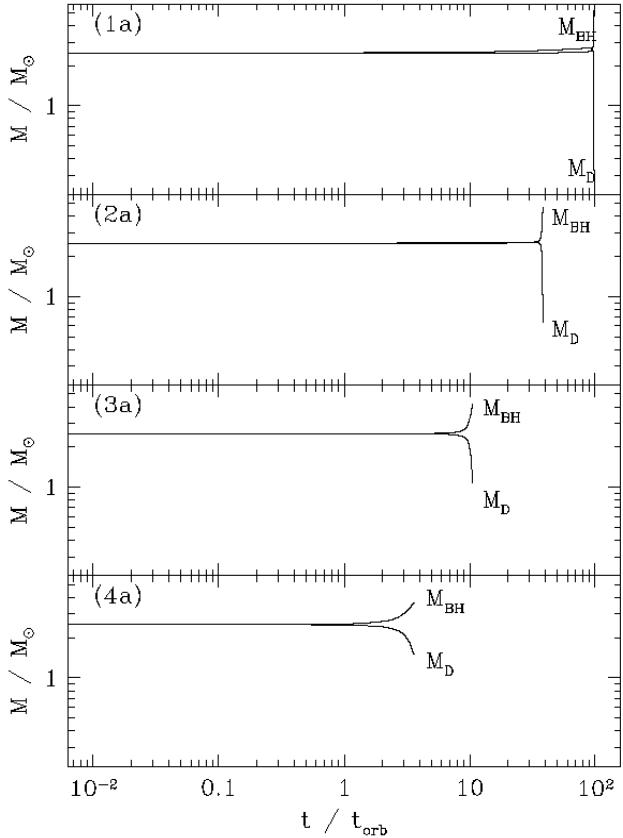,width=\hsize}}
\caption{
Time-evolution of the mass of the black hole and of the mass of the disc for models
(1a) to (4a) listed in Table \ref{tab:ModelParameters}. The sudden appearance of the
instability is noticeable. The increase of the black hole mass is directly associated
with the corresponding decrease in the mass of the disc.}
\label{fig:Instability3}
\end{figure}

\begin{figure}
\centerline{\psfig{file=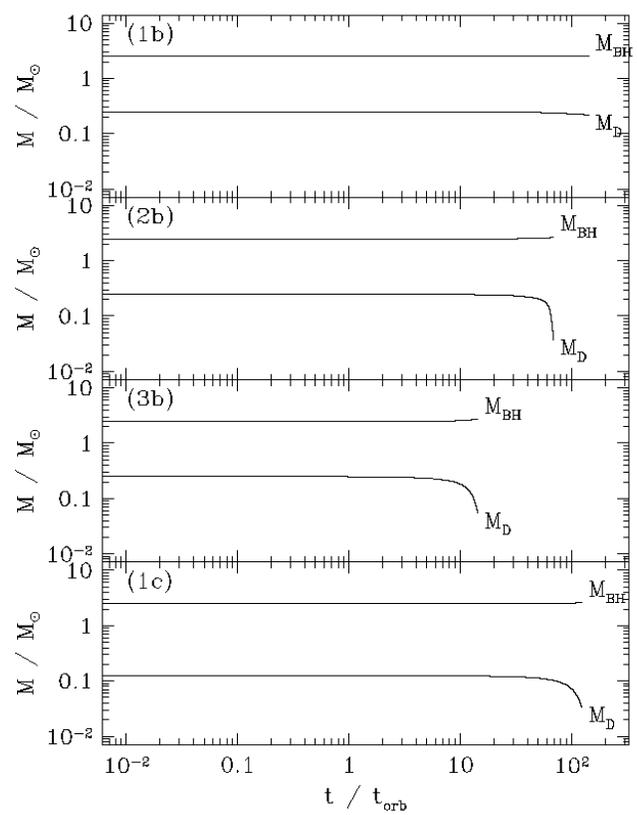,width=\hsize}}
\caption{
Time-evolution of the mass of the black hole and of the mass of the disc for models
(1b) to (3b) and model (1c) of Table \ref{tab:ModelParameters}. Since the mass
of the disc is now much smaller than in models labeled `a' the growth of the
black hole mass is not as clearly visible as in Fig.~\ref{fig:Instability3}. The
sudden decrease of the mass of the disc is however noticeable.}
\label{fig:Instability4}
\end{figure}

The dramatic differences between both series of simulations are depicted in 
Figures \ref{fig:Instability1}-\ref{fig:Instability4}. 
Figures \ref{fig:Instability1} and \ref{fig:Instability2} show the 
time evolution of the mass accretion rate for models (1a) to (4a) and (1b) to 
(3b) plus (1c), respectively. They include, both, the stationary and the unstable cases. 
The overall behaviour found in both figures is similar despite the existing different 
mass ratio between the black hole and the disc, 1 in all curves of Fig. 8 and 0.1 and 0.05 
in those of Fig. 9. The different mass ratio only affects the mass flux and (weakly) the
time in which the instability appears. So, in models labeled `b' and `c' the instability 
takes somewhat more time to appear than in the corresponding models `a' with the same energy 
gap (see below). As a result, models labeled `b' and `c' need more computational time to 
be fully evolved. This fact sets important technical restrictions when trying to evolve models 
with smaller mass ratio for which neglecting the disc self-gravity would be more justified. 
Short-term evolutions of a model with a 0.01 disc-to-hole mass ratio show that the 
instability is indeed present but simply takes longer to grow.

In Figures \ref{fig:Instability1} and \ref{fig:Instability2} we see that 
at early times the evolution of the mass flux for each pair of models is 
exactly the same, irrespective of the increase of the black hole mass being 
taken into account or not. However, whereas the models with a constant black 
hole mass reach a quasi-stationary regime with a constant mass flux, 
the corresponding mass accretion rate for those models with an increasing 
black hole mass goes on increasing after having reached the `stationary' value. 
Furthermore, the time derivative of the mass flux also increases, which implies 
that the mass flux diverges rapidly. 
For all unstable models computed the mass flux is already several orders 
of magnitude larger than the stationary value when the calculation is stopped.
(4 orders of magnitude for Model (1a)!, see Fig.~\ref{fig:Instability1}). 
The divergence found in the mass accretion rate is a clear manifestation of the runaway 
instability at work, which leads ultimately to the complete destruction of the disc.

In Figs. \ref{fig:Instability3} and \ref{fig:Instability4} we plot the time-evolution 
of the black hole mass and the disc mass for the same set of models
displayed in Figs.~\ref{fig:Instability1} and \ref{fig:Instability2}, respectively. 
The sudden loss of the mass of the disc at late times is reflected on the 
corresponding rapid increase of the mass of the black hole. As an example, for 
Model (2a), at $t\sim 40\ t_{\rm orb}$ the black hole has almost doubled its mass 
($M_{\rm BH}\sim 4.7\ \mathrm{M}_{\odot}$) and, correspondingly, the mass of the disc has 
decreased from $2.5\ \mathrm{M}_{\odot}$ to roughly $0.3\ \mathrm{M}_{\odot}$. Note that 
since models `b' and `c' have a much smaller disc mass than models `a' the growth of the 
black hole mass in Fig.~\ref{fig:Instability4} is not as clearly visible as in 
Fig.~\ref{fig:Instability3}.

\begin{figure*}
\begin{center}
\hspace{0.5cm}
\psfig{file=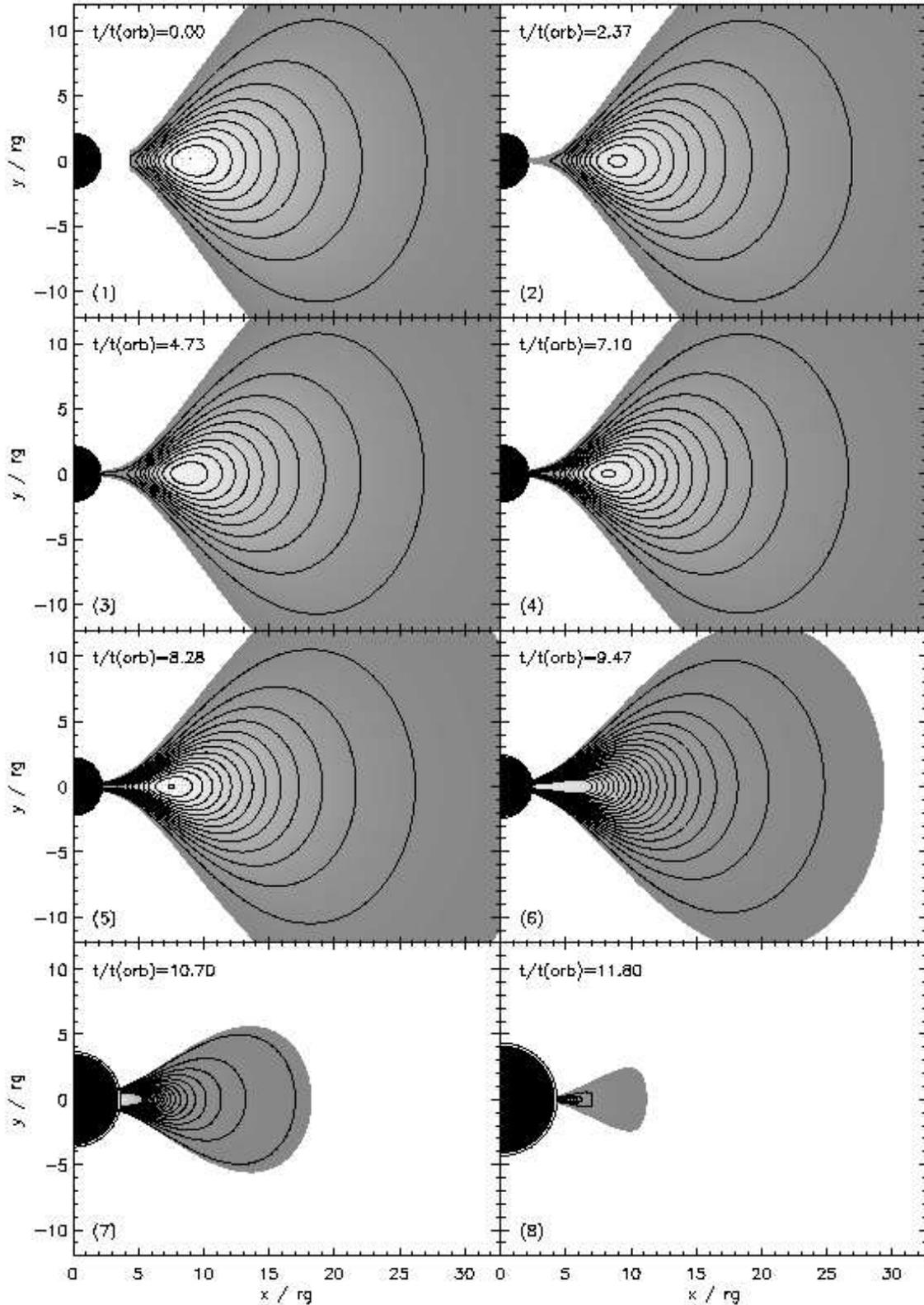,width=15cm}
\end{center}
\vspace*{-0.7cm}
\caption{Time evolution of the unstable model (3a): contour levels of the rest-mass 
density $\rho$ plotted at irregular times from $t=0$ to $t=11.80\ t_\mathrm{orb}$ 
once the disc has almost been entirely destroyed. The density levels are linearly 
spaced with $\Delta\rho=0.1\ \rho_\mathrm{c}^{0}$. The density $\rho_\mathrm{c}^{0}$ 
takes the value at the center of the initial disc (marked by a dot in panel (1)). 
The most exterior contour corresponds to $\rho=0.1\ \rho_\mathrm{max}^{0}$. The entire 
disc is filled in grey color. From panels (1) to (3) the disc is very close to a 
stationary regime and it is almost not evolving. The runaway instability develops 
from panels (4) to (8), most noticeable in the last three panels from time
$t=9.47\ t_\mathrm{orb}$ to $t=11.80\ t_\mathrm{orb}$. The increase in the central 
density and the infall of the disc to the black hole are well visible, as well as 
the associated growth of the black hole.
}
\label{fig:HydroModel3a}
\end{figure*}

The morphology changes that the unstable system undergoes are shown in 
Figure~\ref{fig:HydroModel3a} for a representative case (Model 3a). The 
evolution is qualitatively similar for all models.
In this figure we show eight snapshots of the 
time-evolution from $t=0$ to $t=11.8\ t_{\rm orb}$. The variable plotted 
in the figure is the rest-mass density. The contour levels are linearly 
spaced with $\Delta\rho=0.1\ \rho_\mathrm{c}^{0}$,
where $\rho_\mathrm{c}^{0}$ is the maximum value of the density at the center of the
initial disc. In Fig.~\ref{fig:HydroModel3a} one can clearly follow the transition from a 
quasi-stationary accretion regime (panels (1) to (5)) to the rapid development of the runaway 
instability (panels (6) to (8)). At $t=11.80\ t_\mathrm{orb}$, the disc has almost entirely 
disappeared inside the black hole whose size has noticeably grown. 
From the numerical point of view, and
as already pointed out in Section~\ref{tests} when describing stationary
models, the flow solution remains considerably smooth even though the
evolution is now dynamic. The equatorial plane symmetry is
maintained during the whole evolution with no sign of numerical asymmetries
as well as no vortices appearing inside the disc.

\begin{figure}
\centerline{\psfig{file=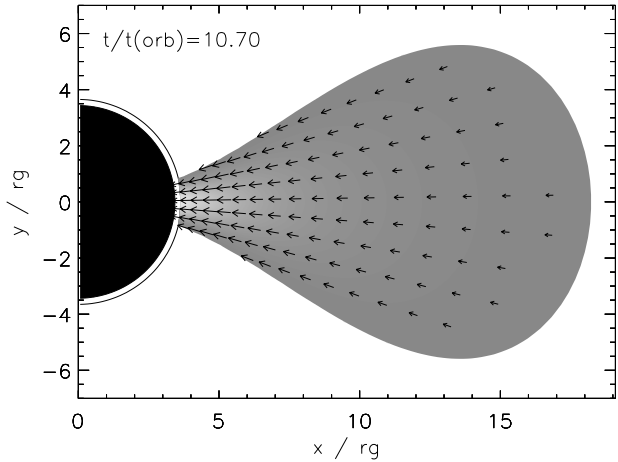,width=8cm}}
\caption{Unstable Model (3a): velocity field at $t=10.70\ t_\mathrm{orb}$ (corresponding 
to snapshot (7) in Fig.~\ref{fig:HydroModel3a}). The arrows are proportional to $v^{i}$ 
and are plotted only in the region where $\rho \ge 0.1\ \rho^{0}_\mathrm{max}$.}
\label{fig:HydroModel3a_velocity}
\end{figure}

\begin{figure}
\centerline{\psfig{file=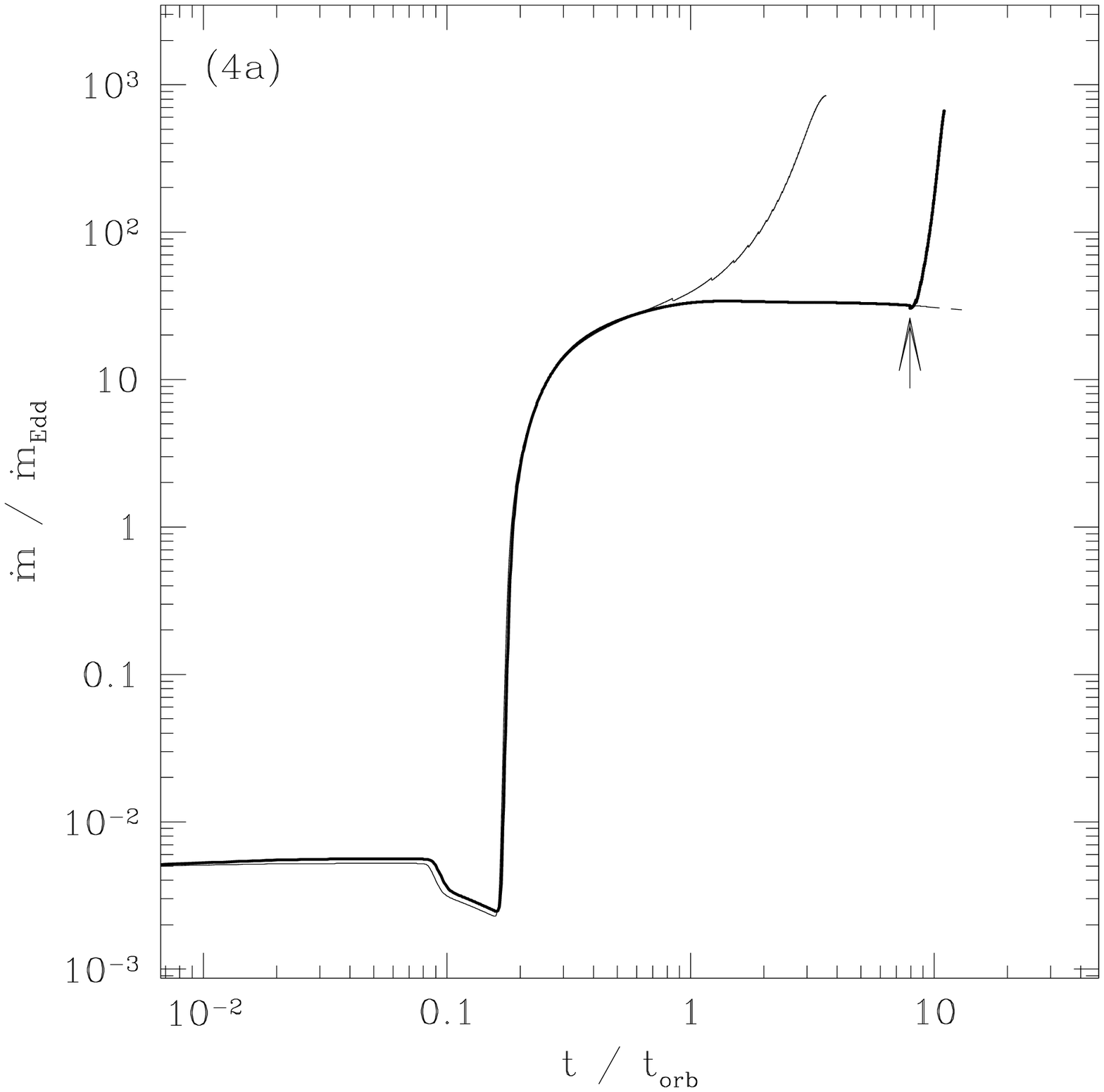,width=\hsize}}
\caption{
Time-evolution of the mass flux for model (4a) of Table \ref{tab:ModelParameters}.
As in Figure \ref{fig:Instability1}, the dashed line corresponds to the case where 
the mass of the black hole is kept constant whereas the thin solid line corresponds 
to the case where the mass of the black hole increases (from $t=0$). In addition, 
the thick solid line corresponds 
to a third case in which the mass of the black hole starts to increase only at time 
$t_{0} = 7.9\ t_\mathrm{orb}$ (indicated by a vertical arrow), i.e. once 
the mass flux has reached its `stationary value'. In this case the instability 
shows up immediately. However, the timescale is the same than in the previous 
run (thin solid line; notice the logarithmic scale for $t$).}
\label{fig:TestTimeScale}
\end{figure}

Correspondingly, Figure \ref{fig:HydroModel3a_velocity} shows the velocity field for 
model (3a) at $t=10.70 t_{\rm orb}$, associated with snapshot (7) in 
Fig.~\ref{fig:HydroModel3a}). This figure shows that the disc is falling radially 
on to the black hole with no signs of vortices and circulation patterns developing.

\begin{figure}
\centerline{\psfig{file=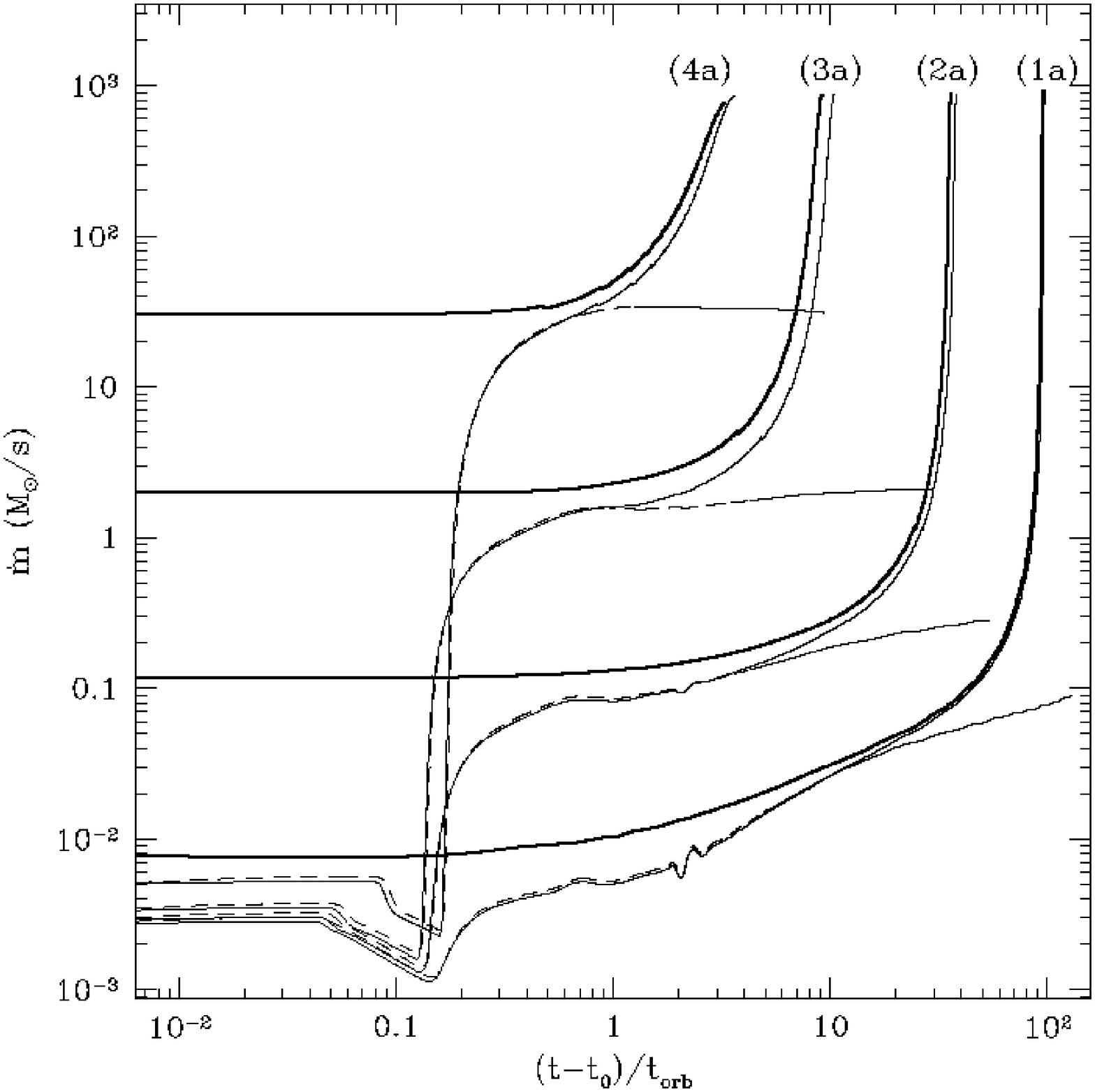,width=\hsize}}
\caption{
Time-evolution of the mass flux as a function of $t-t_{0}$ for models (1a) to (4a) of 
Table \ref{tab:ModelParameters}, where $t_{0}$ is the time at which the mass of the 
black hole starts to increase. As in Figure \ref{fig:Instability1}, the dashed lines 
correspond to the case where the mass of the black hole is kept constant whereas 
the thin solid lines correspond to the case where the mass of the black hole increases 
($t_{0}=0$). The thick solid lines show the evolution when 
the mass of the black hole starts to increase only from time $t_{0}$, equal to 1000, 
200, 24 and 7.9 $t_\mathrm{orb}$ for models (1a) to (4a) respectively, i.e. once 
the mass flux has reached its `stationary value'. In such case, the runaway instability 
appears earlier, the effect being more important for models (1a) and (2a).}
\label{fig:TestTimeScaleSerie1}
\end{figure}

\begin{figure}
\begin{center}
\psfig{file=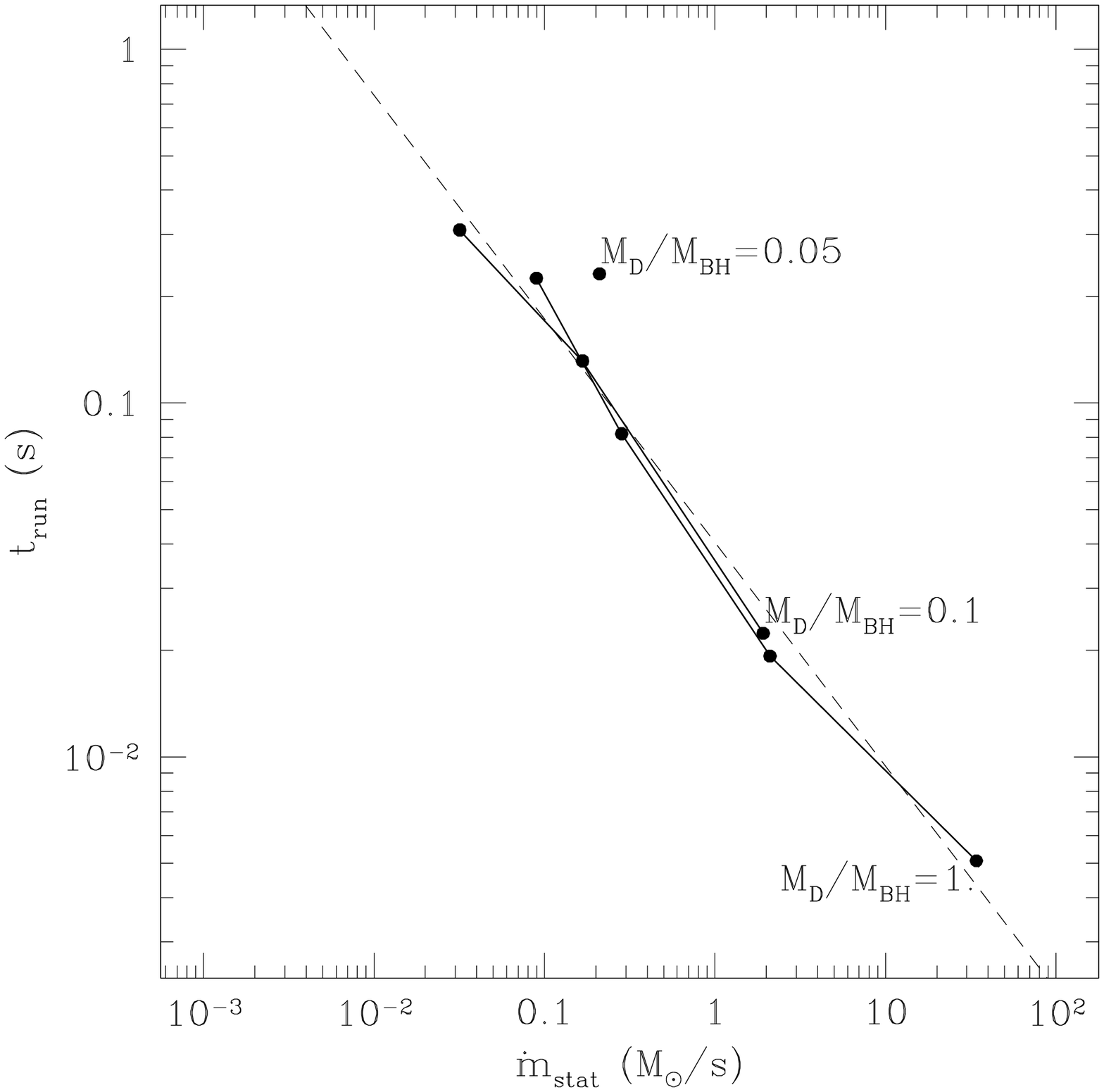,width=\hsize}
\end{center}
\caption{The timescale of the runaway instability $t_\mathrm{run}$ (as defined in Section 
\ref{sec:ResultsUnstable}) is plotted as a function of the mass flux $\dot{m}_\mathrm{stat}$ 
in the `stationary regime' for each of the models listed in Table \ref{tab:ModelParameters}. 
For models (1a) to (4a) ($M_\mathrm{D}/M_\mathrm{BH}=1.$), we use the more accurate estimate 
of $t_\mathrm{run}$ obtained when the mass of the black hole starts to increase once the 
stationary regime has been reached. The dashed line corresponds to the best fit 
$t_\mathrm{run} \propto \dot{m}_\mathrm{stat}^{-\alpha}$ for this series. We find 
$\alpha= 0.9$.}
\label{fig:TimeScale}
\end{figure}

An interesting information which our hydrodynamical simulations provide is the 
timescale of the instability $t_\mathrm{run}$. We estimate this timescale as the 
time it takes for half of the mass of the disc to fall into the hole. 
The values of $t_\mathrm{run}$ obtained for our 8 models are given 
in the last column of Table \ref{tab:ModelParameters}. Such values span 
the interval $\sim 3.8$--$140 t_\mathrm{orb}$, which corresponds to very small 
durations ($\sim 6$--$300\ \mathrm{ms}$). To check the quality of our definition 
of $t_\mathrm{run}$ we have performed the following test: for the four models of 
series `a' we have carried out additional simulations in which the mass of the black 
hole starts to increase only once the stationary regime has been reached at time
$t \simeq t_{0}$. The corresponding evolution of the mass flux in case (4a), for 
which $t_{0}\simeq 7.9\ t_\mathrm{orb}$, is plotted in Fig.~\ref{fig:TestTimeScale}. 
To compare more easily the timescale associated with the runaway instability in the 
two cases, we have plotted in Fig.~\ref{fig:TestTimeScaleSerie1} the evolution of the 
mass flux as a function of $t-t_{0}$. Again we see that in all cases the runaway 
instability appears immediately resulting in the rapid disappearance of the disc. 
However the precise comparison between our usual series of runs (where $t_{0}=0$) and 
the modified ones leads to the conclusion that the timescale of the runaway instability 
is a bit overestimated in the first case. The new values of the timescale for series 
`a' are reported in the last column of Table \ref{tab:ModelParameters}.

In Fig.~\ref{fig:TimeScale} we plot the timescale $t_\mathrm{run}$ as a function of 
the mass flux in the stationary regime $\dot{m}_\mathrm{stat}$.
To be able to derive an empirical law for the timescale of
the instability one should consider a much larger sample of models.
Nevertheless, two clear tendencies can already be extracted from this figure: 
(i) the timescale depends weakly on the disc-to-hole mass ratio; (ii) the runaway 
instability occurs faster when the initial mass flux (stationary value) is larger, 
following approximatively $t_\mathrm{run} \propto \dot{m}^{-\alpha}$. With the value 
of $\alpha=0.9$ obtained for series `a' (where we have used the result of the modified 
runs to get a more accurate estimate of $t_\mathrm{run}$), we can infer that, for all 
cases, the disc is destroyed in a duration never exceeding $1\ \mathrm{s}$ for a large 
range of accretion mass fluxes, $\dot{m}_\mathrm{stat} \ga 10^{-3}\ \mathrm{M_{\odot}/s}$.

\section{Conclusions}
\label{conclusion}

We have presented results from a numerical study of the runaway instability 
of thick discs around black holes. In this study we have carried out a 
comprehensive set of time-dependent simulations aimed at exploring the 
appearance of the instability. In order to do so we have used a fully relativistic,
axisymmetric hydrodynamics code. The general relativistic hydrodynamic equations
have been formulated as a first-order, flux-conservative hyperbolic system and 
solved using a suitable Godunov-type scheme. Among the simplifying conditions we have
assumed a constant angular momentum disc around a Schwarzschild (nonrotating) 
black hole. The self-gravity of the disc has been neglected and the evolution 
of the central black hole has been assumed to be that of a sequence of exact 
Schwarzschild black holes of varying mass.

We have found that by allowing the mass of the black hole to grow the runaway 
instability appears on a dynamical timescale. The mass flux diverges and the disc 
entirely falls into the hole in a few orbital periods (1 $\to$ 100). Therefore,
the appearance of the runaway instability in constant angular momentum discs found in our
simulations is in complete agreement with previous estimates from stationary models 
\citep{abramowicz:83,nishida:96a}. Our simulations provide the first estimation of 
the timescale associated with the runaway instability. For a black hole of 
$2.5\ \mathrm{M_{\odot}}$ and disc-to-hole mass ratios between 1 and 0.05
this timescale never exceeds $1\ \mathrm{s}$ 
for a large range of mass fluxes and it is typically about $50\ \mathrm{ms}$. 
We have found that the dependence of the timescale on the disc-to-hole mass 
ratio is weak and that the runaway instability occurs faster the larger it is 
the initial mass flux (stationary regime) from the disc to the black hole.

We note that our study has been restricted to a polytropic gas, with a particular 
choice of $\kappa$ and $\gamma$ corresponding to a gas of degenerate relativistic 
electrons. We are aware of the over-simplification of such an EoS. However, the work 
of \citet{nishida:96b} has shown that the conclusion of \citet{nishida:96a} (where 
stationary models were built in a fully relativistic computation including the 
self-gravity of the disc) is not modified when using a realistic EoS: constant angular 
momentum discs are unstable. Therefore, we believe that our results would not be 
strongly modified if we were using a more elaborate description of the matter.

To close our investigation we notice that there are four important limitations 
in our study: (i) it is difficult to check the validity of our simple-minded
approach to incorporate the effect of the black hole mass increase;
(ii) the self-gravity of the disc has not been included. Studies based on 
sequences of equilibrium configurations have shown that it favors the 
instability \citep{nishida:96a,masuda:98}; (iii) the rotation of the black hole 
and the possible increase of its spin due to the transfer of angular momentum 
(associated with the transfer of mass) is not yet included in our
current model; (iv) the case of a more realistic distribution of angular momentum 
in the disc (i.e. increasing outwards) has also not been considered yet. 

The first two points in the above list cannot be improved without solving the 
coupled system of Einstein and hydrodynamic equations on black hole spacetimes. Although
important advances in the field of numerical relativity this task is still
challenging. The other items, however, can be more easily improved and work in
this direction will be presented in subsequent investigations. 
In particular, we will present in a forthcoming paper the effect of a non-constant 
angular momentum in the disc. Such a distribution of angular momentum is believed
to suppress the runaway instability according to previous studies in a
stationary framework (e.g.~\citet{daigne:97}). This last point 
- and the very existence of the runaway instability itself - is very important 
in the context of the most discussed scenario for GRBs. In the standard model 
the central engine responsible for the highly energetic emission is a thick disc 
orbiting a stellar mass black hole, with a high accretion mass flux. 
The lifetime of this system must necessarily be larger than a few seconds 
to explain the observed durations of the bursts. Our results show that it would
be absolutely excluded if the runaway instability occurs.

\section*{Acknowledgments}
It is a pleasure to thank Jos\'e Mar\'{\i}a Ib\'a\~nez, Robert Mochkovitch, Luciano
Rezzolla and Olindo Zanotti
for interesting suggestions and a careful reading of the manuscript. 
This research has been mostly supported 
by the Max-Planck-Gesellschaft. J.A.F. acknowledges financial support from a
Marie Curie fellowship from the European Union (HPMF-CT-2001-01172).
F.D. acknowledges financial support from a postdoctoral fellowship from the 
French Spatial Agency (CNES).

\bibliographystyle{mn2e}
\bibliography{paper}
\end{document}